\RequirePackage{lineno}
\documentclass[11pt,titlepage]{article}

\title{Characterization of amylin-induced calcium dysregulation in rat ventricular cardiomyocytes}

\usepackage{xcite}
\usepackage[scientific-notation=true]{siunitx}
\DeclareSIUnit[number-unit-product = {}]
\amp{A}
\DeclareSIUnit[number-unit-product = {}]
\farad{F}
\DeclareSIUnit[number-unit-product = {}]
\molar{M}
\DeclareSIUnit[number-unit-product = {}]
\siemen{S}

\usepackage{xr}
\makeatletter
\newcommand*{\addFileDependency}[1]{
  \typeout{(#1)}
  \@addtofilelist{#1}
  \IfFileExists{#1}{}{\typeout{No file #1.}}
}
\makeatother

\newcommand*{\myexternaldocument}[1]{%
    \externaldocument{#1}%
    \addFileDependency{#1.tex}%
    \addFileDependency{#1.aux}%
}

\myexternaldocument{main_supp}

\usepackage{listings}
\usepackage{chemmacros}
\usepackage{color}
 \definecolor{deepred}{rgb}{0.5,0,0}
 \definecolor{deepgreen}{rgb}{0,0.5,0.2}
 \definecolor{deepblue}{rgb}{0,0,0.65}
 \definecolor{navyblue}{rgb}{0,0.23,0.65}
 \definecolor{brightblue}{rgb}{0.3,0.5,1}
 \definecolor{pink}{rgb}{1.0,0.1,1.0}
 \usepackage{booktabs}
\usepackage{float} 
\usepackage{verbatim}
\usepackage{latexsym}
\usepackage{amsmath}
\usepackage{amsfonts}
\usepackage{amsmath, amsthm, amssymb}
\usepackage{breqn}
\setkeys{breqn}{breakdepth={1}}
\usepackage{euscript}
\usepackage{tabularx}
\usepackage[figuresright]{rotating}

\usepackage[
        style=altlist,
        toc=true,
        ]{glossaries} 

\DeclareGraphicsExtensions{.pdf,.jpg,.eps}
\usepackage[round,numbers,sort&compress]{natbib} 
\newcommand\href[2]{#2 (link)}

\newcommand{\bdsnote}[1]{\mnote{BDS:#1}}

\newenvironment{packed_item}{
\begin{itemize}
  \setlength{\itemsep}{1pt}
  \setlength{\parskip}{0pt}
  \setlength{\parsep}{0pt}
}{\end{itemize}}
\newenvironment{packed_enum}{
\begin{enumerate}
  \setlength{\itemsep}{1pt}
  \setlength{\parskip}{0pt}
  \setlength{\parsep}{0pt}
}{\end{enumerate}}

\newcommand\lbi{\begin{packed_item}}
\newcommand\lei{\end{packed_item}}

\newcommand{\fig}{Fig.~\ref}

\newcommand{\sect}{Sect.~\ref}

\newcommand{\tbl}{Table~\ref}

\newcommand{\eqn}{Eq.~\ref}

\newcommand{\uM}{$\mu M$}

\newcommand{\mynote}[1]{\textcolor{red}{ #1 }}

\newcommand{\codeline}[1]{\lstinline|#1|}
\newcommand{\figsquick}[5]
{
\begin{figure}[ht]
\begin{center}
\includegraphics[width=#5]{#1}
\end{center}
\caption{
\mynote{(figshort:#2)}
#3\label{figshort:#2}
}
\end{figure}
}

\newcommand{\figbig}[4]
{
  \figsquick{#1}{#2}{#3}{#4}
  {6in}
}

\newcommand{\clm}{Cl$^{-}$}
\newcommand{\catwo}{Ca$^{2+}$}

\newcommand{\nap}{Na$^{+}$}
\newcommand{\kp}{K$^{+}$}
\newcommand{\lip}{Li$^{+}$}

\newcommand{\BAR}{$\beta$-adrenergic}

\renewcommand\sl{sarcolemmal}
\newcommand\etal{\textit{et al.}}

\newacronym[description={Ca2+/calmodulin-dependent protein kinase II}]
{camkii}{CaMKII}{Ca2+/calmodulin-dependent protein kinase II}
\newcommand\camkii{\gls{camkii}}
\newacronym[description={Protein kinase A}]
{PKA}{PKA}{Protein kinase A}
\newacronym[description={gain-of-function}]
{GOF}{GOF}{gain-of-function}
\newacronym[description={loss-of-function}]
{LOF}{LOF}{loss-of-function}
\newacronym[description={Ryanodine receptor}]
{ryr}{RyR}{ryanodine receptor}
\newacronym[description={root mean squared fluctuations}]
{RMSF}{RMSF}{root mean squared fluctuations}
\newacronym[description={root mean squared deviations}]
{RMSD}{RMSD}{root mean squared deviations}
\newacronym[description={nuclear magnetic resonance}]
{NMR}{NMR}{nuclear magnetic resonance}
\newacronym[description={heart failure}]
{HF}{HF}{heart failure}
\newacronym[description={knock-out}]
{KO}{KO}{knock-out}
\newacronym[description={G-protein coupled receptor}]
{GPCR}{GPCR}{G-protein coupled receptor}
\newacronym[description={non-junctional sarcoplasmic reticulum}]
{NSR}{NSR}{non-junctional sarcoplasmic reticulum}
\newacronym[description={sub-sarcolemmal space}]
{SSL}{SSL}{sub-sarcolemmal space}
\newacronym[description={junctional sarcoplasmic reticulum}]
{JSR}{JSR}{junctional sarcoplasmic reticulum}
\newacronym[description={Cardiac ventricular myocyte}]
{CVM}{CVM}{cardiac ventricular myocyte}
\newacronym[description={principal components analysis}]
{PCA}{PCA}{principal components analysis}
\newacronym[description={Brownian dynamics}]
{BD}{BD}{Brownian dynamics}
\newacronym[description={thermodynamic integration}]
{TI}{TI}{thermodynamic integration}
\newacronym[description={replica exchange molecular dynamics}]
{REMD}{REMD}{replica exchange molecular dynamics}
\newacronym[description={molecular dynamics}]
{MD}{MD}{molecular dynamics}
\newacronym[description={accelerated molecular dynamics}]
{AMD}{AMD}{accelerated molecular dynamics}
\newacronym[description={phospholamban}]
{PLB}{PLB}{phospholamban}
\newacronym[description={Diabetic cardiac myopathy}]
{DCM}{DCM}{diabetic cardiac myopathy}
\newacronym[description={Congestive heart failure}]
{CHF}{CHF}{congestive heart failure}
\newacronym[description={Ligand-binding domain}]
{LBD}{LBD}{ligand-binding domain}
\newacronym{SB}{SB}{Shannon-Bers}
\newacronym{SBM}{SBM}{Shannon-Bers-Morroti }
\newacronym[plural=DADs]{dad}{DAD}{delayed after-depolarization}
\newacronym{ucd}{UCD}{Type II diabetes UCD model}

\newacronym{apd}{APD}{action potential duration}
\newcommand\apd{\gls{apd}}
\newacronym{hip}{HIP}{human amylin transgenic}
\newcommand\hip{\gls{hip}}
\newacronym{amy}{+Amylin}{acute amylin-exposed rats}
\newcommand\amy{\gls{amy}}
\newacronym{sr}{SR}{sarcoplasmic reticulum}
\newcommand\sr{\gls{sr}}
\newacronym{SL}{SL}{sarcolemma}
\newcommand\SL{\gls{SL}}
\newacronym{ec}{EC}{excitation-contraction}
\newcommand\ec{\gls{ec}}
\newacronym{bar}{$\beta$AR}{$\beta$-adrenergic receptor} 
\newacronym{ga}{GA}{genetic algorithm}
\newcommand\ga{\gls{ga}}
\newcommand\betaar{\gls{bar}} 
\newacronym{ap}{AP}{action potential}
\newcommand\ap{\gls{ap}}
\newacronym{tnc}{TnC}{Troponin C}
\newcommand\tnc{\gls{tnc}}
\newacronym{cam}{CaM}{calmodulin}
\newcommand\cam{\gls{cam}}
\newacronym{ncx}{NCX}{\nap/\catwo\ exchanger}
\newcommand\ncx{\gls{ncx}}
\newacronym{cicr}{CICR}{\catwo-induced \catwo\ release}

\newacronym{nka}{NKA}{\nap/\kp\ atpase}
\newcommand\nka{\gls{nka}}
\newacronym{hf}{HF}{heart failure}

\newacronym{pka}{PKA}{protein kinase A}
\newcommand\pka{\gls{pka}}
\newacronym{lcc}{LCC}{L-type calcium channel}
\newcommand\lcc{LCC}
\newacronym{nfat}{NFAT}{Nuclear factor of activated T-cells}
\newcommand\nfat{\gls{nfat}}
\newacronym{hdac}{HDAC}{Histone deacetylases}
\newcommand\hdac{\gls{hdac}}
\newacronym{serca}{SERCA}{Sarcoplasmic/endoplasmic reticulum calcium ATPase}
\newcommand\serca{\gls{serca}}
\newacronym[plural={transverse tubules}]{tt}{TT}{transverse tubule}
\newacronym[plural=ODEs]{ode}{ODE}{ordinary differential equations}
\newacronym[plural=CRUs]{cru}{CRU}{Ca$^{2+}$ release unit}
\newacronym{sda}{SDA}{State Decomposition Analysis}
  \newcounter{mnote}
  \newcommand{\mnote}[1]{\addtocounter{mnote}{1}
    \ensuremath{{}^{\bullet\arabic{mnote}}}
    \marginpar{\footnotesize\em\color{red}\ensuremath{\bullet\arabic{mnote}}#1}}
  \let\oldmarginpar\marginpar
    \renewcommand\marginpar[1]{\-\oldmarginpar[\raggedleft\footnotesize
#1]%
    {\raggedright\footnotesize #1}}

\newcommand\ODE{\gls{ode}}

\newcommand{\iclca}{$i_{\mbox{Cl(Ca)}}$}
\newcommand{\iclb}{$i_{\mbox{Clb}}$}
\newcommand\ica{$i_{\mbox{Ca}}$}
\newcommand\icap{$i_{\mbox{Cap}}$}
\newcommand\icab{$i_{\mbox{CaB}}$}
\newcommand\inab{$i_{\mbox{NaB}}$}
\newcommand\inak{$i_{\mbox{NaK}}$}
\newcommand\itof{$i_{\mbox{tof}}$} 
\newcommand\itos{$i_{\mbox{tos}}$} 
\newcommand\ikr{$i_{\mbox{Kr}}$}  
\newcommand\iks{$i_{\mbox{Ks}}$}   
\newcommand\iki{$i_{\mbox{K1}}$}  
\newcommand\ikp{$i_{\mbox{Kp}}$} 
\newcommand\ikur{$i_{\mbox{kur}}$} 
\newcommand\iss{$i_{\mbox{ss}}$} 
\newcommand\ilcc{$i_{\mbox{CaL}}$}
\newcommand\incx{$i_{\mbox{NaCa}}$}
\newcommand\ina{$i_{\mbox{Na}}$}
\newcommand\slna{$SL_{\mbox{Na}}$}
\newcommand\jctna{$jct1_{\mbox{Na}}$}
\newcommand\ryr{\gls{ryr}}

\newcommand\sda{\gls{sda}}
\newcommand\sbm{\gls{SBM}}
\renewcommand\sb{\gls{SB}}

\renewcommand\mynote[1]{}
\renewcommand\mnote[1]{}
\newcommand\sdnote[1]{}

\author{
Bradley D. Stewart$^1$ \and
Caitlin E. Scott$^1$ \and
Thomas P. McCoy$^2$ \and     
Guo Yin$^3$ \and
Florin Despa$^3$ \and
Sanda Despa$^3*$ \and
Peter M. Kekenes-Huskey$^1*$}   
\date{
  $^1$ Department of Chemistry, University of Kentucky, Lexington, KY, USA 40506\\
  $^2$ Department of Family \& Community Nursing, University of North Carolina - Greensburo, Greensburo, NC, USA\\
  $^3$ Department of Pharmacology and Nutritional Sciences, University of Kentucky, Lexington, KY, USA 40506\\
  \today
}

\begin{document}
\maketitle



\abstract{
Hyperamylinemia, a condition characterized by above-normal blood levels of the pancreas-derived peptide amylin, accompanies obesity and precedes type II diabetes. 
Human amylin oligomerizes easily and amylin oligomers deposit in the pancreas \citep{Westermark2011}, brain \citep{Verma2016}, and heart \citep{Liu2016},
where they have been associated with calcium dysregulation.
In the heart, accumulating evidence suggests that human amylin oligomers form modestly cation-selective \citep{Sciacca2012,Mirzabekov1996}, voltage-dependent ion channels that embed in the cell sarcolemma (SL).
The oligomers increase membrane conductance in a dose-dependent manner \citep{Mirzabekov1996}, which
is correlated with elevated cytosolic \catwo. 
These effects can be reversed by pharmacologically disrupting amylin oligomerization \citep{DESPA2014}. 
These findings motivated our core hypothesis that non-selective inward \catwo\ conduction afforded by human amylin oligomers increase cytosolic and \sr\ \catwo\ load, which thereby magnifies intracellular \catwo\ transients. 
\label{intro:questions}
Questions remain however regarding the mechanism of amylin-induced \catwo\ dysregulation, including 
whether enhanced SL \catwo\ influx is sufficient to elevate cytosolic \catwo\ load \citep{Despa2012},
and if so, how might amplified \catwo\  transients perturb \catwo-dependent cardiac pathways. 
To investigate these questions, we modified a computational model of cardiomyocytes \catwo\ signaling to reflect experimentally-measured changes in SL membrane permeation and decreased \serca\ function stemming from acute and transgenic human amylin peptide exposure.
With this model, we confirmed the hypothesis that increasing SL permeation alone was sufficient to enhance \catwo\ transient amplitudes without recruitment of prominent SL-bound \catwo\ transporters, such as the L-type \catwo.
Our model indicated that amplified cytosolic transients are driven by increased \catwo\ loading of the sarcoplasmic reticulum and may contribute to the \catwo-dependent activation of calmodulin.
These findings suggest that increased membrane permeation induced by deposition of amylin oligomers contributes to \catwo\ dysregulation in pre-diabetes.
}

\newpage
\section*{Introduction}


Amylin, a 3.9 kilodalton peptide produced by the pancreatic $\beta$ cells \citep{Cooper1987}, is secreted along with insulin into the blood stream \citep{Pieber1993}. 
Increased circulation of human amylin peptide preceding the onset of type II diabetes has been correlated with amylin deposits in the heart \citep{Haataja2008}.
These deposits have been shown to induce  diastolic dysfunction \citep{Despa2012}, hypertrophy, and dilation \citep{DESPA2014}. 
While the amylin peptide in humans is amyloidogenic, that is, it polymerizes into amyloid-like fibrils, rodents secret a non-amyloidogenic form of amylin that does not accumulate in cells or tissue.
However, rodents expressing human amylin in the pancreatic $\beta$ cells develop late onset type-2 diabetes \citep{Westermark2011,Haataja2008}.
While studies correlating human amylin depositing with the onset of pathological states typical of diabetic cardiomyopathy \citep{Bugger2014} are beginning to emerge \citep{Liu2016}, molecular mechanisms linking amylin insult with cellular dysfunction remain poorly understood.
Gaining momentum, however, is the notion that amylin oligomerization in cardiac tissue may disrupt normal calcium homeostasis \citep{Despa2012}, stemming from its modestly cation-selective conductance properties. \citep{Sciacca2012,Mirzabekov1996,JosephASchauerte2010}.
While this conductance is small relative to predominant \SL\ \catwo\ currents including the \lcc\ and \ncx, it nevertheless exhibits largely unexplained effects on perturbing intracellular \catwo\ signals and recruiting \catwo-dependent pathways associated with pathological, hypertrophic remodeling \citep{Roderick2007}. 

To motivate the interrelationship between amylin and potential \catwo\ dysregulation in the heart, we first summarize key aspects of cardiac \catwo\ signaling.
In the healthy heart, the \catwo-dependent \ec\ coupling cycle begins with a depolarizing \ap\ that modulates \SL\ \catwo\ fluxes, including the \acrfull{lcc}, \acrfull{ncx}, and sarcolemmal \catwo\ leak \citep{Bers2001}.
\catwo\ entry via LCC and NCX triggers \citep{Litwin1998}  \acrfull{sr} \catwo\ release via \glsplural{ryr}, leading to a rapid increase in intracellular \catwo (\catwo\ transient) that ultimately activates and regulates competent myocyte contraction \citep{Bers2001}.
The cycle completes as SR \catwo\ uptake via the \acrfull{serca}, as well sarcolemmal \catwo\ extrusion via NCX and the sarcolemmal \catwo\ ATPase, collectively restore diastolic \catwo\ levels.
Recently, we reported that this process was perturbed in rats transgenic for human amylin (\hip), as well as in isolated cardiomyocytes acutely exposed to the peptide (\amy) \citep{Despa2012}.
In particular, both rat models exhibited larger cytosolic \catwo\ transients and faster rates of sarcolemmal \catwo\ leak than control. 
Furthermore, we found that in \hip\ rats, SERCA function was impaired and hypertrophic remodeling associated \nfat/\hdac pathways were activated; both properties are strongly associated with the progression toward heart failure \cite{Roderick2007}.
In this study, therefore, we seek to clarify mechanisms by which human amylin-induced \sl\ \catwo\ leak leads to \catwo\ dysregulation in pre-diabetes and ultimately the activation of hypertrophic remodeling pathways.

Cardiac computational models are particularly well-suited for exploring intracellular mechanisms of \catwo\ signaling and their dysregulation in cardiac tissue \citep{Gong2017,Hake2014,Winslow2012, Li2010}.
We extended one such model, the Shannon-Bers model of ventricular myocyte \catwo\ dynamics \citep{Shannon2004}, to unravel the influence of amylin in the \hip\ phenotype.
Specifically, the revised model reflects our experimentally-measured changes in SL membrane \catwo\ permeation as well as decreased \serca\ function consistent with acutely-exposed and transgenic human amylin rats \citep{Despa2012}.
We find that increased \catwo\ background leak conductance via amylin was sufficient to reproduce enhanced \catwo\ transients  previously measured in \hip\ rats \citep{Despa2012}.
These simulations implicate increased \sr\ loading as the primary mechanism of increasing \catwo\ transient amplitude for the amylin phenotypes, which in turn elevates cytosolic \catwo\ load.
Finally, we show higher propensities for \cam\ activation under conditions of elevated diastolic \catwo, which we speculate may trigger the \cam-dependent \nfat\ remodeling pathway. 
These findings lead to our hypothesized model of amylin-induced \catwo dysregulation summarized in \fig{figshort:hypoth}. 

\figbig{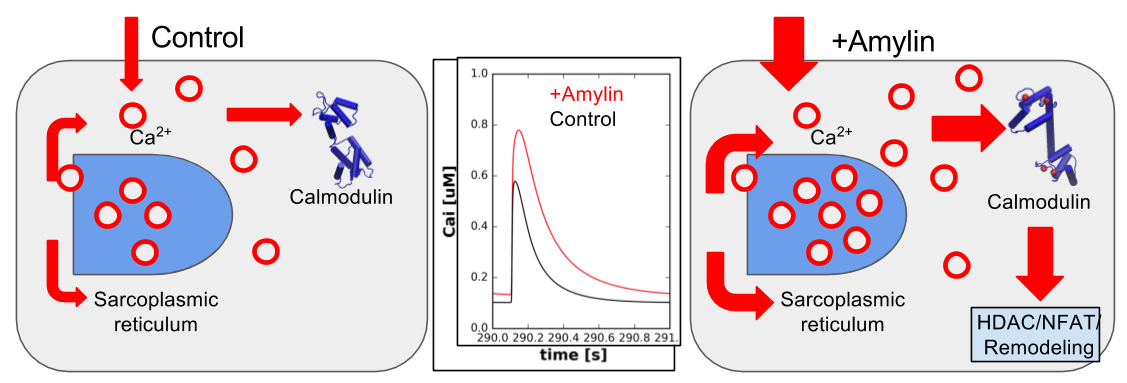}
{hypoth}
{Hypothesized model.
Increased sarcolemmal \catwo\ in \acrfull{amy} increases sarcoplasmic reticulum \catwo\ loading, amplifies intracellular \catwo\ transients and increases the \catwo-bound state of proteins including \acrfull{cam} (PDB codes 1DBM and 3CLN). 
}
{N/A}


\section*{Materials and Methods}
\subsection*{Experimental animals}
N=12 Sprague-Dawley rats were used in this study. 
All animal experiments were performed
conform to the NIH guide for the care and use of laboratory animals and were approved by the
Institutional Animal Care and Use Committee at University of Kentucky. \sdnote{do we need a number for the approval?}
Ventricular myocytes
were isolated by perfusion with collagenase on a gravity-driven Langendorff apparatus \citep{Despa2012}.

\subsection*{Measurements of \catwo\ transients and sarcolemmal \nap/\catwo\ leak}

Myocytes were plated on laminin-coated coverslips, mounted on the stage of a fluorescence microscope and loaded with Fluo4-AM (10 $\mu$mol/L, for 25 min). 
\catwo transients were elicited by stimulation with external electrodes at a frequency of 1 Hz.
The passive trans-sarcolemmal \catwo\ leak was measured as the initial rate of \catwo\ decline upon reducing external \catwo from 1 to 0 mM. 
In these experiments, \catwo fluxes to and from the \sr\ were blocked by pre-treating the cells with 10 $\mu$M thapsigargin for 10 min whereas the \ncx\ and sarcolemmal \catwo - ATPase were abolished by using 0 \nap/0 \catwo solution (\nap\ replaced with \lip) and adding 20 $\mu$M carboxyeosin, respectively.
The outward sarcolemmal \catwo\ leak was measured in the absence and presence of the membrane sealant poloxamer 188, which is a surfactant that stabilizes lipid bilayers and thus protects against amylin-induced sarcolemmal damage. 

\nap\ influx was measured as the initial rate of the increase in intracellular \nap\ concentration ([\nap]$_i$) immediately following \nka\ pump inhibition with 10 mM ouabain.
As described previously \citep{Despa2002}, [\nap]$_i$ was measured using the fluorescent indicator SBFI (TefLabs).
The SBFI ratio was calibrated at the end of each experiment using divalent-free solutions with 0, 10, or 20 mmol/L of extracellular \nap\ in the presence of 10 $\mu$mol/L gramicidin and 100 $\mu$mol/L strophanthidin. 

\subsection*{L-type \catwo\ current measurement}
L-type \catwo current (\ica) was measured under voltage-clamp in whole cell configuration. 
\ica\ was determined as the nifedipine-sensitive current recorded during depolarization steps from -40 mV (where the cell was held for 50 ms to inactivate \nap\ channels), to -35 to +60 mV.
The patch-pipette was filled with a solution containing (in mM) 125 Cs-methanesulfonate, 16.5 TEA-Cl, 1 \ch{MgCl2}, 10 EGTA, 3.9 \ch{CaCl2} , 5 Hepes, and 5 Mg-ATP (pH=7.2).
The external solution contained (in mM) 150 NMDG, 1 \ch{CaCl2} , 5 4-aminopyridine, 1 \ch{MgCl2}, 10 Hepes, and 10 glucose (pH=7.4).

\subsection*{Simulation and analysis protocols}
\subsubsection*{Summary of Shannon-Bers-Morotti rat \catwo\ handling model} 
To examine the relationship between increased \sl\ \catwo\ entry and elevated \catwo\ transients reported in rats \citep{Despa2012}, we adapted a rabbit ventricular myocyte model of \catwo\ signaling  to reflect handling terms specific to mice and rats.
Our choice of a mouse model was based on the initial lack of rat-specific \catwo\ handling models available in the literature and the similar rates of \catwo\ relaxation via \serca, \ncx, and other minor contributions (\sl) shared by rat and mice (92, 8 and 1\%, versus 90.3, 9.2 and 0.5\%, respectively)  shared by both species \citep{Li1998,Bassani1994}. 
Mouse-specific parameter and potassium current changes were introduced into the Shannon-Bers rabbit cardiomyocyte \catwo\ 
model \citep{Shannon2004} according to Morotti \etal. \citep{Morotti2014} (summarized in Supplement).
The resulting model is hereafter referred to as the \sbm\ model. 
Model equations, 'state' names, current names and initial conditions are provided in the supplement. 
As noted in \citep{Morotti2014}, four predominant changes in potassium channels were included: 
1) the transient outward potassium current expression for rabbits was replaced with fast component (\itof) for mice,
2) the slowing activating delayed rectifier current was substituted with a slowly inactivating delayed rectifier current (\iks),
3) a non-inactivating potassium steady-state current (\iss) was added
4) the inward rectifier potassium current (\iki) was reduced.
Other distinctions between the two species are the elevated intracellular sodium load and sodium ion current in murine versus rabbit species, which we optimized to match experimental data collected in this study.
In \fig{figshort:rabbitvsmouse}-\fig{figshort:othercurrents}, we compare metrics such as \catwo\ transients, action potentials, potassium currents and prominent \nap/\catwo\ currents for rabbits 
versus mice, for which we report 
excellent agreement with data from Morotti \etal. \citep{Morotti2014}.

\label{meth:leak}
To reflect increased SL leak due to amylin in \hip\ rats, we modulated the Shannon-Bers leak model described by 
\begin{equation}
\label{eqn:leak}
I_{iBK} = Fx_{iBk}\;G_{iBk}(V-E_{i})
\end{equation}
where $Fx$ represents leak density, $G_c$ is the max conductance for \catwo\ $i$, $V$ is voltage and $E_{i}$ is the Nernst potential of ion $i$. 
In rats acutely exposed to amylin (\amy), amylin oligomer deposition was correlated with a roughly 70\% increased rate of \catwo\ leak in hypotonic solution (see Figure 3D of Despa \etal \citep{Despa2012}).
We accordingly increased $G_c$ for \catwo\ by a commensurate amount (see \tbl{tbl:params}) to reflect this observation.
Although amylin pores exhibit poor cation selectivity \citep{Mirzabekov1996}, we maintained $G_c$ for \nap\ at baseline values, given that we observed no detectable change in \nap\ load (\fig{figshort:naexp}).
Furthermore, though we assume that the increased \catwo\ influx scaled with applied voltage ($V$), given the short duration of the action potential this approximation did not significantly affect the model.
The magnitude of enhanced SL \catwo\ leak, $I_{Cab}$, is depicted in \fig{figshort:Cabsetc}.
Lastly, we fit our model outputs \nap\ transients and $\tau$ values to mimic the rat using a \ga\ (detailed in the Supplement) that varied the parameters of the \nka\ current and the \serca\ Vmax values respectively.
This fitting was necessary to capture \catwo\ and \nap\ transients reported in the \hip, \amy\ and an 'activated LCC' condition that mimics enhanced \SL\ \catwo\ leak.
We compare our predicted normalized \catwo\ transients to control data in \fig{figshort:} and report excellent agreement. 
For our \amy\ configuration, the \ga\ increased \nka\ Vmax by 14\% 
to maintain normal \nap\ levels (see \tbl{tbl:params}), which concurs with a study indicating agonized \nka\ function in skeletal muscle \citep{Clausen2004}.
For the human amylin transgenic model (\hip), it was observed that \catwo\ transient decay time increased by nearly 30\% 
relative to control \citep{Despa2012}), for which the \ga\ determined a 
reduction in \serca\ Vmax by 47\% was necessary.
Lastly, we introduced an 'LCC' configuration for which \catwo\ permeation is increased, to elucidate potential differences between \catwo-entry via non-selective leak as opposed to via \catwo-selective channels (see \tbl{tbl:params}).

\figbig{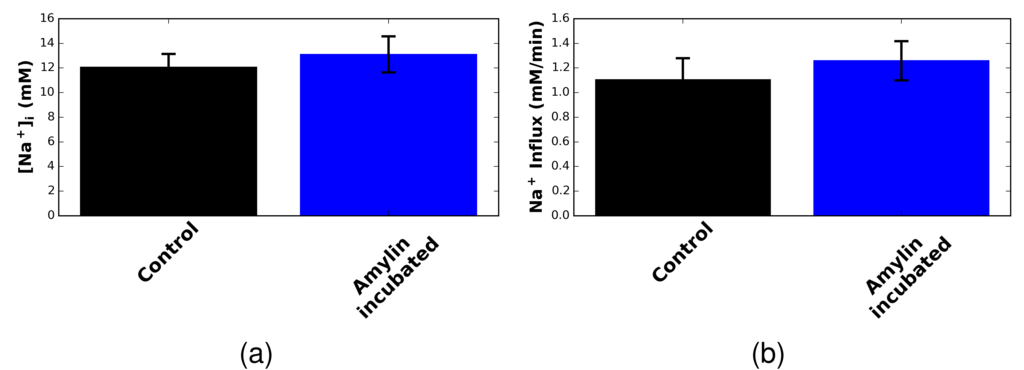}
{naexp}
{Experimental measurements of sodium load and influx in control and myocytes} 
{morottiModel_daisychain.ipynb}

\subsubsection*{Numerical model of \catwo\ handling} 
The Shannon-Bers cellML  model was converted into a Python module via the  Generalized  ODE  Translator gotran (https://bitbucket.org/johanhake/gotran).
The mouse-specific alterations summarized in the previous section were implemented into the resulting module.
In our numerical experiments, the \sbm\ model was numerically integrated  by the scipy function \textsc{odeint}, which utilizes the LSODA algorithm for stiff ordinary differential equations \citep{PETZOLD1983}.
The numerical model was integrated using a timestep of 0.1 ms for a total simulation time of up to 5 minutes.  
These simulations provide as output the time-dependent values of the \sbm\ 'states', such as intracellular \catwo\ load or the action potential, as well as 'currents' that include major \catwo, \nap, \kp, and \clm-conducting proteins. 
Model fitting proceeded by a genetic algorithm  (reviewed in \citep{Srinivas1994}) that iteratively improved parameter values, such as LCC \catwo\ conductance, membrane leak, and NKA conductance over several generations of 'progeny'  (\fig{figshort:algVerify}).
Experimentally-measured outputs, such as \catwo\ transient decay time and amplitude, were measured for each of the progeny; those that reduced output error relative to the experimentally-measured equivalent with stored for future generations (see \sect{supp:fitting}) for more details).
To validate our implementation, we present comparisons of action potentials, intracellular \catwo\ and \nap\ transients, as well as ionic currents for rabbit versus murine cardiac ventricular 
myocytes in \sect{supp:morottimodel}, and we report good agreement.
Steady-state action potentials and intracellular \catwo\ oscillations were generally observed within 6000 ms. 
We additionally stimulated the model at several frequencies ranging from 0.25 to 2.0 Hz, with the reference frequency at 1.0 Hz, subject to default (control) parameters as case-specific values listed in Methods.
Sensitivity of \catwo\ transients to \sl\ leak rates and Vmax for \nka\ and \serca\ were probed as described in \sect{supp:sensanaly}.
All data processing was performed using scipy and the ipython notebook; source code will be provided at 
\textsc{https://bitbucket.org/huskeypm/wholecell}. 

\subsubsection*{Analyses}
To examine potential mechanisms that link increased \SL\ \catwo\ permeation to \sr-loading and elevated \catwo\ transients, we present a simple method, \sda, that monitors and identifies prominent changes in key 'state' variables (including the action potential, \sr-load, channel gate probabilities among others) as well as ion channel currents, relative to control conditions.
The key benefit of this approach is the automated identification of modulated EC coupling components that can motivate model refinements and additional experiments.
The \sda\ method consists of the following steps: 
1) numerically solve the time-dependent \ODE s governing all components (the state variables) of the \ec\ coupling model for trial and control parameter configurations 
2) 'score' the time-dependent state values according to metrics like amplitude  
3) calculate percent differences between trial and control state variable scores
4) rank order states by either the percent difference with a reference state or by the amplitudes in the reference.

\newpage

\section*{Results}
\subsection*{Effects of human amylin on intracellular \catwo\ transients in rat cardiac ventricular myocytes}  
The accumulation of human amylin aggregates in rat cardiomyocyte \SL\ was previously correlated with increased rates of \sl\ \catwo\ conduction and amplified \catwo\ transient amplitudes \citep{Despa2012}.
Increased \catwo\ \sl\ conduction was originally attributed to permeation across the bilayer, as opposed to direct modulation of ion channels, based on observations of an amylin dose-dependent 
outward \catwo\ leak \citep{Despa2012}.
This led to the hypothesis that sarcolemma-localized human amylin oligomers have the primary effect of increasing the inward \catwo\ leak, although the mechanism linking \catwo\ leak and transient amplitudes was not established.
We first validate this hypothesis by measuring the effect of human amylin (50 $\mu $M; 2 hours incubation) on \SL\ \catwo\ leak and \catwo\ transient amplitude in the absence and in the presence of poloxamer 188 (P188), a surfactant that stabilizes lipid bilayers through hydrophobic interactions \citep{Collins2007}. 
As reported previously, human amylin significantly increased both passive \SL\ \catwo\ leak  and \catwo\ transient amplitude (\fig{figshort:sealant}A). 
When amylin was applied in the presence of P188 (50 $\mu$M), however, \SL\ \catwo\ leak and transient amplitudes were statistically comparable to control (\fig{figshort:sealant}). 
Similar behavior was observed upon co-incubation with epoxyeicosatrienoic acids (14,15-EETs; 5 $\mu$M), which have anti-aggregation effects and reduce amylin deposition at the \SL\ \citep{DESPA2014}. 
These results support the hypothesis that amylin primarily influences myocyte \catwo\ cycling through poration of the sarcolemma.

\figbig{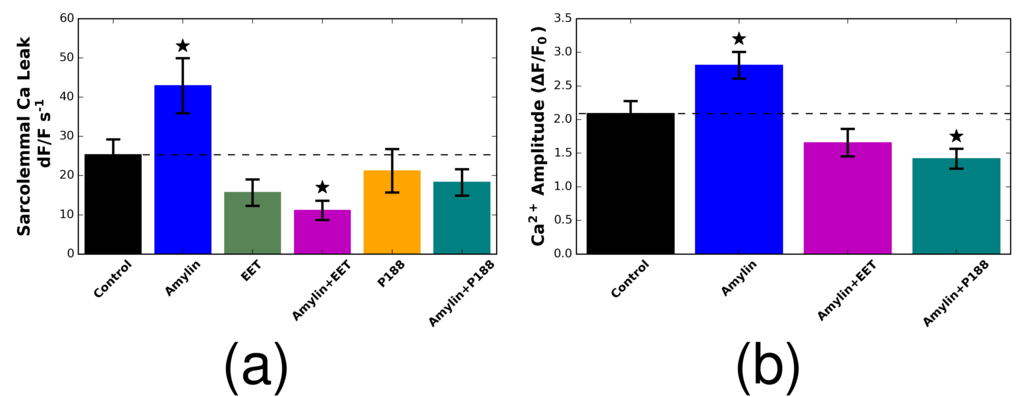}
{sealant}
{Sarcolemmal \catwo\ leak and intracellular \catwo\ transient amplitude.
a) Outward \sl\ \catwo\ leaks are reported for control and amylin-incubated rats. 
Significantly higher leak rates were found for amylin-incubated rats relative to control. 
Introduction of membrane sealants EET and P188 maintained \catwo\ leak rates at levels comparable to control conditions. 
b) \catwo\ transients under analogous conditions are elevated for amylin-incubated rats, while control and sealant-exposed myocytes exhibit equivalent amplitudes}
{morottiModel_daisychain.ipynb}

To investigate how membrane poration via human amylin leads to amplified \catwo\ transients, we numerically solved the \sbm\ whole-cell model at 1 Hz pacing under control conditions and with enhanced \SL\ leak (\amy).
Our simulations confirmed that \catwo\ transients for the \amy\ configuration  were 46\% higher than control (\fig{figshort:incrLeak}), consistent with experiment (\fig{figshort:sealant}b and Fig, 3C of \citep{Despa2012}). 
We further modeled the pre-diabetic (\hip) rats  examined in \citep{Despa2012} by assuming increased \SL\ leak and decreased SERCA function.
Similar to \amy\, the \hip\  model predicted elevated \catwo\ transient amplitudes (36\% relative to control), although they were somewhat attenuated compared to the \amy\ conditions.
In contrast to the \amy\ configuration, however, the \hip\ model presented \mynote{XXX} 27\% slower diastolic relaxation \bdsnote{Fix me (see also \fig{figshort:expvssimNorm})} and a \mynote{XXX} 92 \% increase in  diastolic intracellular \catwo\ load relative to control, as would be expected with reduced SERCA function \cite{Pereira2006}.
\bdsnote{Fix the XXXs and missing ref}
We further note that the enhancement of \catwo\ transient amplitudes for \amy/\hip\ rats relative to control diminished with increased pacing (up to 2 Hz), in accordance with experimental findings (see \fig{figshort:ratfigure5D}). 
\catwo\ transient relaxation rates remained unchanged over this range, as our model does not currently include factors governing frequency dependent acceleration of relaxation, such as the involvement of \camkii \citep{DeSantiago2002}.

A distinctive feature of murine species is the dominant role of the SR in managing \catwo\ homeostasis, with nearly 90\% of the 
intracellular \catwo\ transient originating from \sr\ \citep{Bers2001}.
In contrast, in higher species, \sl\ derived \catwo\ plays a significantly larger role; in rabbits, for instance, inward \sl\ \catwo\ currents account for roughly 40\% of the intracellular \catwo\ transient \citep{Bers2001}. 
As a proof of principle, we augmented the original \sb\ formulation of cardiac \catwo\ cycling in rabbits \citep{Shannon2004} with increased \catwo\ leak.
In \fig{figshort:rabbitamyNoNKAleak300pct} 
, we demonstrate similar trends of increased cytosolic and \sr\ \catwo\ load under conditions of increased \sl\ \catwo\ leak. \bdsnote{Add value for rabbits, put into \tbl{tbl:params}}).  

\figbig{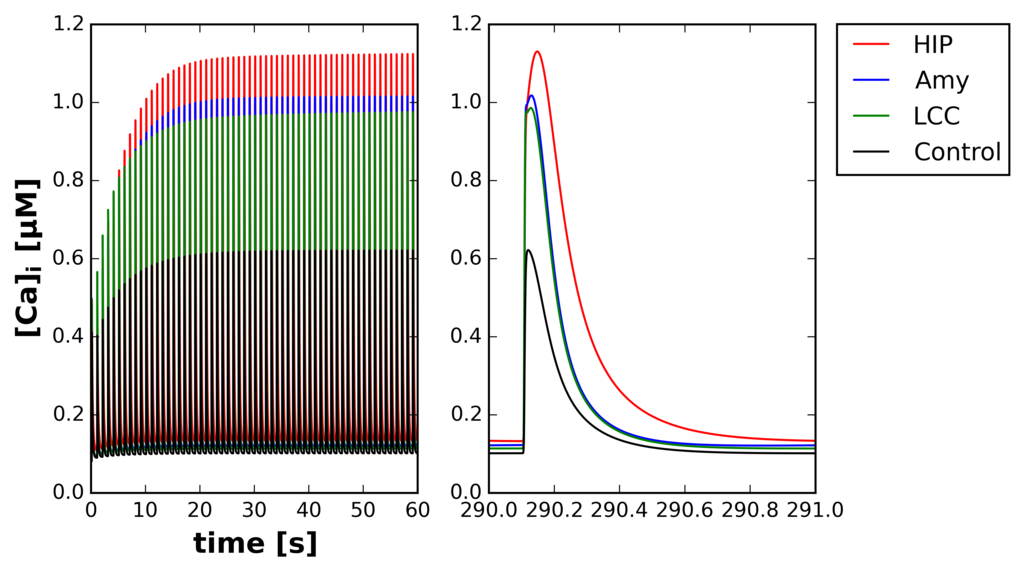}
{incrLeak}
{
Intracellular \catwo\ transients (concentration versus time) predicted using the \acrfull{SBM} \catwo\ cycling model following 300s of 1.0 Hz pacing.
Transients are reported for model conditions representing control (black),
\acrfull{amy} (blue),
\acrfull{hip} (red) and \acrfull{lcc} (green).
Comparisons between simulation and experiment from 0.5 through 2 Hz pacing are shown in \fig{figshort:ratfigure5D}
}
{morottiModel_daisychain.ipynb}

\subsection*{Effects of acute amylin-induced modulation of sarcolemmal ion handling}  
\subsubsection*{Cytosolic and sarcoplasmic reticulum \catwo\ load} 
While background \sl\ \catwo\ leak is evidently enhanced for \amy\ and \hip, the corresponding \catwo\ current over a single beat does not contribute significantly to the total cytosol \catwo\ content.
Hence, the leak alone is insufficient to directly account for the observed increase in \catwo\ amplitude for the amylin models on a beat-to-beat basis.
Rather, our data indicate that the \catwo\ transients required nearly 20 seconds of pacing to reach steady state (\fig{figshort:incrLeak}), which suggests that \catwo\ transient amplification occurs through an alternative mechanism.
Since the majority of the \catwo\ released on a beat-to-beat basis originates in the SR \citep{Bers2001}, we hypothesized that the increased intracellular \catwo\ transient amplitudes for the amylin rats stemmed from elevated SR \catwo\ loading owing to increased \sl\ \catwo\ leak. 
For this scenario, we would expect that \catwo\ transient amplitudes should scale proportionally with \SL\ leak rates. 
Therefore, we examined how the control model responded to variations SL \catwo\ leak (Amylin Leak \%), as well in SERCA function.
These effects are summarized in \fig{figshort:3dchartSERCA}a-c, for which we report predicted cytosolic \catwo\ transients (a., $\Delta Cai$), SR \catwo\ transients (b., $\Delta Ca\_SR$) and diastolic SR \catwo\ loads (c., max $Ca\_SR$).  
These data strongly indicate that the SR \catwo\ load is positively correlated with increasing \sl\ \catwo\ leak and to a lesser extent, SERCA function.
More importantly, the increased \sl\ \catwo\ leak assumed for 
\amy\ and \hip\ relative to control largely accounted for the elevated \catwo\ transients and \sr\ load. 

In other words, SERCA appeared to play a minor role in tuning the \catwo\ transient in our amylin model, as the reduced SERCA Vmax for \hip\ relative to \amy\ maintained enhanced, albeit modestly reduced, \catwo\ transients and load. 
Instead, SERCA  control the extent to which altered \sl\ \catwo\ leak modulates \catwo\ transient amplitude. 
This was most apparent as pacing rates were varied from 0.5 to 2 Hz in our model, which essentially determined the time during which SERCA could recover \sr\ \catwo\ load following a release event.
Specifically, our model predicted that amylin-induced \catwo\ transient enhancement diminished with increased pacing and nearly approached control transient amplitudes at 2 Hz  (see \fig{figshort:ratfigure5D}).
Further, the decline in transient amplitude with pacing was faster for \hip\ relative to \amy, which expectedly suggests that amylin's inotropic effects are at least partially modulated by the  efficiency of SERCA \catwo\ handling. 

\figbig{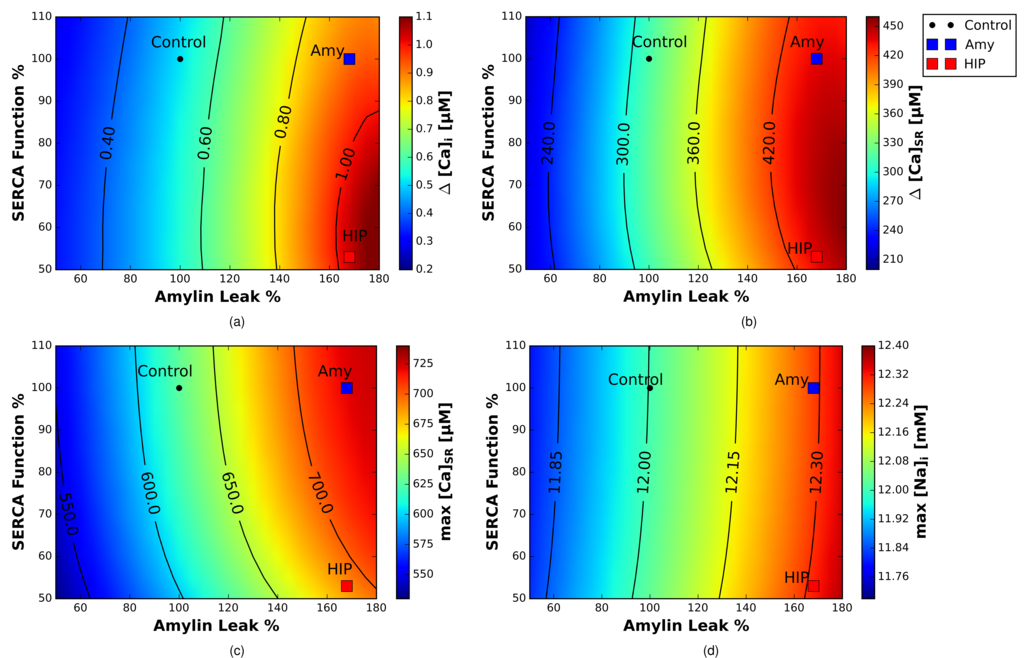}
{3dchartSERCA}
{Predicted \catwo\ transients and loads as a function of SERCA Vmax activity (\% of control) and SL \catwo\ leak (\% of control). 
a) intracellular \catwo, b) SR \catwo\ transient c) maximum SR 
\catwo\ load and d) sodium load. A black point is representative of the Control case, a blue square is representative of the Amy case, and a red square is representative of the HIP case. 
}
{morottiModel_daisychain.ipynb}

\subsubsection*{Maintenance of \nap\ load in \amy\ and \hip\ myocytes}
To determine whether amylin induced appreciable changes in cardiomyocyte \nap\ handling, we measured \nap\ load and influx in control and amylin-incubated myocytes following the inhibition of the \nka\ pump.
\nka\ is a sarcolemma-bound ATPase that extrudes \nap\ by exchanging the cation with extracellular \kp, thus its inhibition would be expected to demonstrate any differences in \nap\ load and influx due to \amy.
We found that both \nap\ handling metrics were indistinguishable between the control and \amy\ cells when the \nka\ was allowed to compensate for the \catwo\ leak (\fig{figshort:nka}c). 
However, when we assumed \nka\ activity in our \amy\ model was identical to control, we found that the intracellular \nap\ increased by 0.3 mM 
(\fig{figshort:nka}c), though the predicted difference in \nap\ load was likely below the limits of experimental detection.
The increased \nap\ load appeared to arise due to higher \ncx\ exchange rates (\fig{figshort:comparativecurrentsAmplitude}) brought about by the elevated diastolic \catwo\ load for \amy. 
In order to maintain \nap\ load at control levels for the \amy\ model, our fitting procedure revealed that the \nka\ current should be increased by 14\%.
Interestingly, it has been reported \citep{Clausen2004} that rat soleus muscle exposed to 10 \uM\ amylin increased Rb cation uptake by 24\% relative to control \citep{Clausen2004}, which is commensurate with our predicted Vmax for maintaining cytosolic \nap\ load. 

To further elucidate the potential contribution of \nka\ exchange to \catwo\ and \nap\ homeostasis, we present in \fig{figshort:3dchartNKA} cytosolic and \sr\ \catwo\ transient amplitudes as well as \nap\ load as a function of \sl\ leak rates and \nka\ activity. 
In \fig{figshort:3dchartNKA}d we confirm that \nap\ load decreases with increasing \nka\ Vmax and increases with \sl\ \catwo\ leak. 
Our model assumes amylin does not change \sl\ \nap\ leak relative to control, therefore we attribute the positive correlation between \nap\ load and \sl\ \catwo\ leak to NCX exchange activity.
Specifically, as cytosolic \catwo\ load increases with \sl\ leak rates, \ncx\ exchange of cytosolic \catwo\ with extracellular \nap\ would contribute to increased intracellular \nap. 
Analogously, as increased \nka\ activity depletes cytosolic \nap, \catwo\ influx via the \ncx\ reverse mode would be expected to decrease and thereby ultimately reduce intracellular \catwo.
Our simulated data reflect these trends for several metrics of \catwo\ transients in \fig{figshort:3dchartNKA}a-c, which we discuss further in the Supplement.

\subsubsection*{Ion channel activity and \catwo\ handling}
It was expected that amylin-driven increases in cytosolic and SR \catwo\ loading would culminate in the modulation of multiple downstream \catwo-dependent signaling pathways \citep{Despa2002}.
In this regard, we leveraged the computational model to systematically probe the response of its outputs, such as the activity of various \catwo\ handling components, to changes in model inputs including SL \catwo\ leak. 
Accordingly, we depict in \fig{figshort:comparativecurrentsAmplitude} relative changes in all ion channel amplitudes described in the \sbm\ model for the \amy\ and \hip\ configurations, ranked by their absolute magnitudes. 
These data expectedly reflect increased sarcolemmal \catwo\ leak (\icab) for \amy\ and \hip, as we assumed increased leak conductance parameters for both cases. 
Interestingly, \ina\ was predicted to increase for both cases relative to control, which in principle could influence the \ap\ upstroke velocity \citep{Lowe2008}.
However, in \fig{figshort:amy}d we demonstrate that the \ap\ waveform is largely unchanged in the amylin cases relative to control, thus the predicted effects on \ina\ amplitude appear to be of little consequence.
Beyond these currents, increased \SL\ leak had opposing effects on the currents for the \amy\ and \hip\ data.
For \amy, for instance, we observed enhanced \sr\ release and uptake amplitudes ($j_{\mbox{relSR}}$, $j_{\mbox{pumpSR}}$ and $j_{\mbox{leakSR}}$ in \fig{figshort:comparativecurrentsAmplitude})
\bdsnote{Embed SR crap in figshort:comparativeSR into main body 'current' image DONE } that are expected to contribute to larger cytosolic \catwo\ transients. 
For \hip, we found modestly higher \incx\ and \icap\ relative to control and \amy, which reflects a redistribution of sarcolemmal \catwo\ extrusion versus \sr\ \catwo\ uptake. 
Similar redistributions are known to occur when SERCA function is reduced \cite{Bovo2014}. 

\figbig{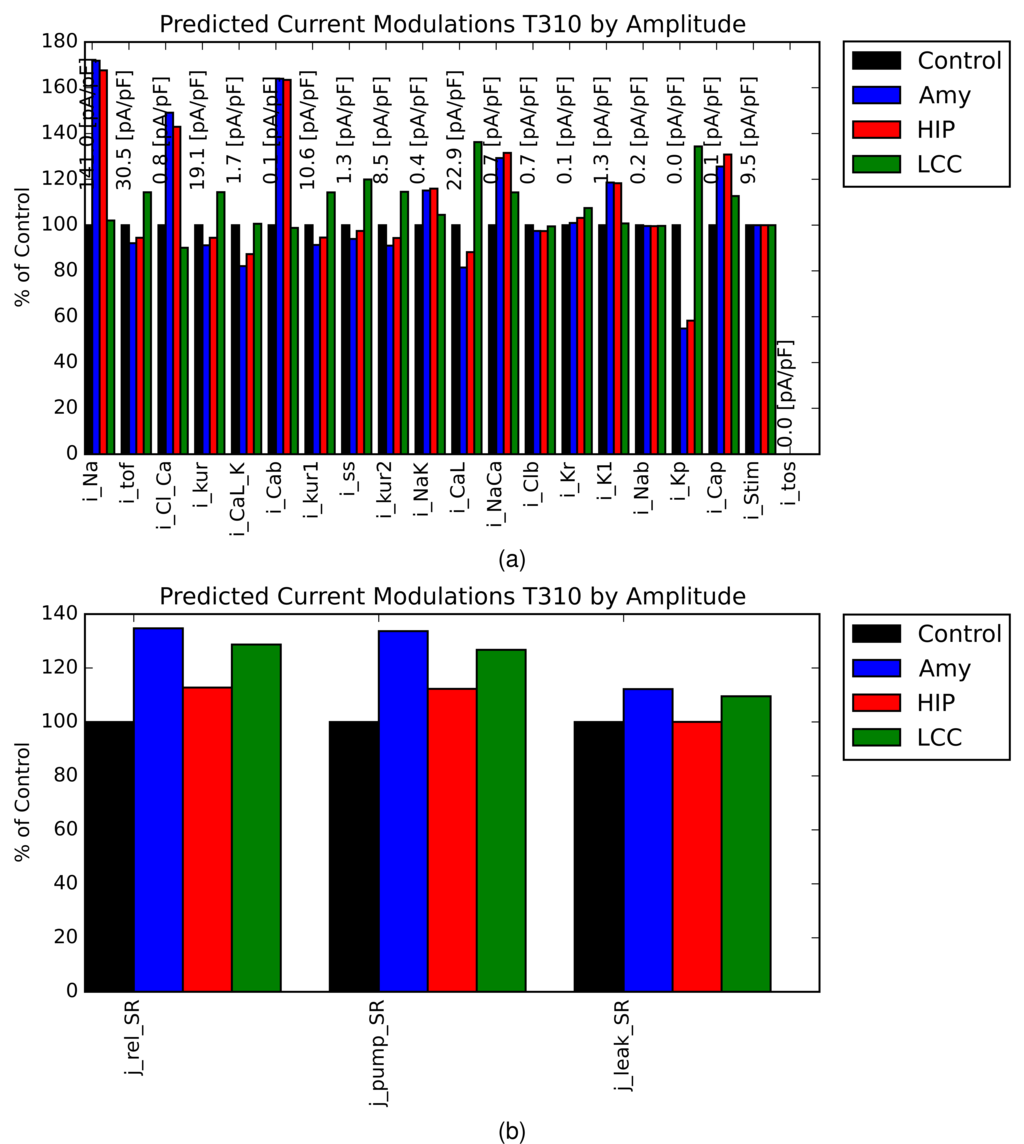}
{comparativecurrentsAmplitude}
{Percent change in \sbm-predicted ion current amplitudes for \amy (blue), \hip (red) and increased LCC conductance (green, see \fig{figshort:lcc}) configurations relative to control (black, 
normalized to 100\%) for temperature 310K sorted by current amplitude change.
A list of current labels is provided in the supplement \tbl{tbl:modelterms}, as well as the currents' normalized values 
(\fig{figshort:comparativecurrents})}
{morottiModel_daisychain.ipynb}

In \fig{figshort:comparativestates} we depict the relative change in activity for the top twenty modulated model 'states' upon increasing SL \catwo\ leak.
Unique to \amy\ were nearly 2.5- and 1.5-fold increases in the inactive (I) and open (O) states of the Ryanodine receptor model \citep{Stern1999,Stern1997} relative to control,  which is consistent with  elevated dyadic junction \catwo\ that acts to both promote and terminate \ryr\ opening.
More importantly, the greater RyR open probability translates to an increased rate of SR \catwo\ release and commensurate increase in cytosolic \catwo\ transients. 
Apparent to both \amy\ and \hip\ conditions are 30-75\% increases in states representing intracellular \catwo\ and \catwo-bound buffers, including \cam, \tnc, and myosin, which can be expected with \catwo\ loading. 

\subsection*{Comparison with non-amylin-induced increases in sarcolemmal \catwo-specific currents and sarcoplasmic reticulum \catwo\ handling}
Our main hypothesis was that the \amy\ phenotype is primarily driven by non-specific SL \catwo\ leak.
This mechanism would be in contrast to direct modulation of \catwo\ conducting channels, as has been demonstrated for LCC and TRPV4 in neurons \citep{Zhang2017}.
To investigate these hypotheses, we fit the LCC conductance to reproduce the cytosolic \catwo\ transients exhibited for \amy.
The fitting procedure yielded an increased PCa value relative to control (180\%) \bdsnote{which val DONE} that in turn increased peak \ilcc. 
These conditions, which we refer to as the \lcc\ configuration, were found to present many of the same trends observed for the \amy\ conditions, including increased intracellular \sr\ \catwo\ transients and SR load (see \fig{figshort:lcc}).
\mnote{BDS: CES writes: Figures in the supplement aren't in order.}
Our analyses in \fig{figshort:comparativecurrents} revealed some differences in \sl\ channel currents for \lcc\ relative \amy. 
Firstly, the data reflect the model assumptions of higher \catwo\ leak for the \amy\ case and larger magnitude \ilcc\ for \lcc.
More importantly, \amy\ and \lcc\ were found to have different effects on \ina, as the former indicated an amplified sodium channel current, while the \ina\ for  \lcc\ was similar to control.
Conversely, the most prominent \kp\ channel currents (\itof, \ikur, and \iki) were for the most part moderately enhanced for \lcc, compared to modest suppression of those current for \amy. 
Despite these opposing effects on \ina\ and \kp\ currents, there were negligible differences in the \ap\ relative to control  (see \fig{figshort:Cabsetc}i).

\clearpage
\newpage
\section*{Discussion}

\subsection*{Shannon-Bers-Morotti myocyte model}
We revised the Shannon-Bers model of rabbit ventricular myocyte \catwo\ cycling \citep{Shannon2004}
to reflect \catwo\ handling in murine species, as a close approximation to the human amylin transgenic/amylin-exposed rats used in \citep{Despa2012}.
The predominant changes implemented in our model  primarily entailed increasing the rates of \sr\ \catwo\ uptake and release to mirror the larger role of SR \catwo\ handling in murine relative to higher order animals, as well as modulating potassium channel current profiles. 
The \sbm\ model captured key distinguishing features of murine cardiomyocyte \catwo\ handling,
including shorter \ap\ and \catwo\ transient duration relative to rabbit, as well as a greater role of \catwo\ release and uptake via the \sr, as opposed to \ncx \citep{Bers2002}.
When we included \sl\ \catwo\  leak data from Despa \etal. \cite{Despa2012} appropriate for the \amy\ and \hip\ phenotypes in rats, as well as reduced SERCA \catwo\ uptake rates for \hip, the computational model reproduced the altered \catwo\ transient amplitudes across a broad range of pacing intervals.
With this model, we conclude that 
\lbi
\item increased rates of \catwo\ influx through the sarcolemma, for instance as a result of amylin-induced membrane poration, promotes the amplification of cytosolic \catwo\ transients.  
\item the increase in \catwo\ transient amplitude arises due to greater \sr\ \catwo\ load relative to control
\item elevated cytosolic \catwo\ load stemming from higher rates of \sl\ \catwo\ influx (\amy), and especially when SERCA function is reduced (\hip), significantly increases the proportion of \catwo-bound proteins.  Of these proteins, \cam\ activation in particular may trigger remodeling via the calcineurin/\nfat\ pathway \cite{Wilkins2004}(see \fig{figshort:hypoth}).
\item the concerted relationship between amylin-induced increased 
sarcolemmal \catwo\ leak, intracellular \catwo\ transients, and \sr\ loading gives rise to similar \catwo\ transient amplification in the Shannon-Bers model of EC coupling in rabbit \citep{Shannon2004}, which suggests similar mechanisms of dysregulation in pre-diabetes may be manifest in higher order mammals.
\lei

\subsection*{Enhanced SL \catwo\ fluxes are sufficient to elevate cytosolic \catwo\ load in absence of altered SR \catwo\ handling}
Recently, it was established that pre-diabetic rats transgenic for human amylin peptide  presented a high density of oligomerized amylin deposits in ventricular tissue \citep{Despa2012}. 
Cells containing these deposits were additionally found to have greater sarcolemmal \catwo\ leak rates and amplified \catwo\ transients. 
These effects on \sl\ \catwo\ conductance and transient amplitudes were recapitulated in isolated myocytes that were incubated with human amylin, which suggested that the phenotypical changes likely precede any significant changes in protein expression that might otherwise produce similar effects. 
Further, disruption of amylin oligomers via increasing eicosanoid serum levels \citep{DESPA2014}  and the application of membrane sealant P188 (\fig{figshort:sealant}) were both found to restore normal \catwo\ handling.
These experiments together firmly establish the link between oligomer-induced membrane poration and \catwo\ dysregulation. 
Similarly, in our implementation of the Morotti-Shannon-Bers \catwo\ cycling model, we found that amplified \catwo\ transients could be induced solely by increasing the \sl\ \catwo\ conductance parameter (see \eqn{eqn:leak}).

The enhancement of intracellular \catwo\ transient amplitudes by amylin bears similarity to agonism of the \sl\ \catwo\ channels \lcc\ and P2X.
It is well-established, for instance, that activation of \lcc\ via \betaar\ agonists promote larger \catwo\ transients that are accompanied by elevated SR \catwo\ load \citep{Bers2001}.
Further, P2X receptor activation has comparable effects on 
\catwo\ transients and SR load \citep{Shen2006}, albeit without the multifarious changes in \catwo\ handling associated with \betaar\ stimulation.
While we defer the topic of SR load to later in the Discussion, our simulations present strong evidence that increased inward \sl\ \catwo\ alone is sufficient to explain amylin dose-dependent 
effects on \catwo\ transients in \citep{Despa2012}.

For pacing intervals at 1 Hz and greater, our predictions of the control \catwo\ transient using the \sbm\ model (see \fig{figshort:ratfigure5D}) follow a neutral transient amplitude/frequency relationship, as is frequently exhibited in 
mice \citep{Antoons2002} and the Despa \etal 
rat control data \citep{Despa2012}.
Further, the computational model captures the negative \catwo\ transient relationships with pacing frequency reflected in the Despa \etal 
\hip\ rat data, including the diminishing difference in transient amplitude relative to control at 2 Hz pacing.
The decline in transient amplitude for \hip\ can be ascribed to the inability to maintain elevated SR load as pacing increases, given the reduced SERCA activity evident for these rats \cite{Despa2012}. 
Our data also reflect a negative transient amplitude/frequency relationship for the \amy\ conditions, which may arise because the model does not reflect phosphorylation-dependent effects on 
relaxation, including \camkii\ activation \citep{Bassani1995}.
Nevertheless, given that our model captures the predominant changes in \catwo\ handling exhibited in \amy\ and \hip\ pre-diabetic rats \cite{Despa2012} chiefly through modulating \sl\ \catwo\ leak, our simulations support the hypothesis that increased SL \catwo\ entry alone (without recruiting cation-specific channels like LCC) promotes the development of enhanced \catwo\ transients (see \fig{figshort:hypoth}). 

\subsubsection*{Contributions of SR loading to amylin phenotype }
We demonstrated in \fig{figshort:diffCaSRConc} a positive correlation of increasing \catwo\ SL leak rates with elevated SR \catwo\ loading and transients, respectively,  with preserved \serca\ function.
This configuration is analogous to the \amy\ conditions assumed in this study.
Therefore, the predicted amplification of the cytosolic \catwo\ transients appears to be driven by \catwo-loading of the SR, which in turn affords greater RyR \catwo\ flux per release event. 
We note that diastolic SR \catwo\ load was modestly increased by approximately 10\% relative to control under the \amy\ conditions.
The increased SR load appeared to be of little consequence, as steady-state behavior was maintained through several minutes of simulated pacing without evidence of \glsplural{dad}.
These results concur with those of Campos \etal,
for which computational studies of rabbit ventricular myocytes indicated considerable tolerance to SR \catwo\ overload before abnormal \ap\ behavior was evident \citep{Campos2015}.
Further, our hypothesis is congruent with a study examining triggering of the SL \catwo\ channel P2X4, which was found to yield both elevated \catwo\ transients and SR \catwo load \citep{Shen2007}\bdsnote{give percentages from this paper. PKH: Do you mean our paper we are writing or the Shen207 paper?}.

An interesting finding from our simulations, is that both \amy\ and \hip\ rats presented amplified \catwo\ transients, despite the latter of which having predicted diastolic SR \catwo\ loads commensurate with the control (see \fig{figshort:hip}).
The notion that diastolic SR \catwo\ loads are comparable for \hip\ and control has precedent, as insignificant changes in SR load relative to control were reported in \citep{Despa2012}.
We speculate that the higher diastolic cytosolic \catwo\ exhibited in \hip\ may amplify \ryr\ release (\fig{figshort:Cabsetc}) via \catwo-induced \catwo\ release \cite{Zahradnikova1999}, which would ultimately yield larger \catwo\ transients despite unchanged \sr\ \catwo\ load.

\subsection*{Implications of elevated cytosolic \catwo\ load}
An interesting consequence of elevated \catwo\ transients and in the case of \hip, increased diastolic \catwo\ load, is the potential for activating \catwo-dependent pathways that are normally quiescent during normal \catwo\ handling. 
We observed in \fig{figshort:comparativestates} for instance, that greater levels of \catwo-bound \cam\ and \tnc\ are  evident relative to control. 
Under normal conditions, \catwo\ activation of \tnc\ is the critical substrate for force development in contractile tissue \citep{Gordon2000}, while \cam\ in part regulates normal force-frequency relationships and responses to \BAR\ stimulation  \citep{Maier2002}.
However, it is also implicated in the activation of pathways associated with remodeling and failure \citep{Roderick2007}. 
In particular, activation of the \cam-regulated CaMKII is attributed to cardiac remodeling via the \hdac\ pathway 
Concurrently, activation of the phosphatase calcineurin via \cam\ is known to promote transcriptional changes by way of \nfat\ activation \citep{Heineke2012}, which together contribute to the hypertrophic response to dysregulated \catwo\ handling \citep{Bers2008}. 
Indeed, in pre-diabetic \hip\ there was evidence that CaMKII-HDAC and calciuneurin-NFAT remodeling were simultaneously activated \citep{Despa2012}.
In this regard, while the increased \catwo\ transients stemming from amylin oligomerization may initially have beneficial inotropic effects, activation of \cam\ and its dependent hypertrophic pathways may contribute to cardiac decline. 

\subsection*{Limitations}
Our model was based on a rather modest set of changes in \catwo, \nap\ and \kp\ handling to a rabbit ventricular cardiomyocyte formulation. 
Further refinement of rat electrophysiology  \citep{Demir2004,Hintz2002} and implementation of a recent rat 
/catwo/ handling model \cite{Gattoni2016}, could provide improved predictive power for our model of amylin-induced dysregulation. 
In the greater context of diabetes, it is likely that the \catwo\ dysregulation and subsequent activation of \camkii\ sets forth a cascade of maladaptive events that drive heart failure.
As such, our simulation results could be improved by including 
the impact of altered \pka\ and \camkii\ activity on \catwo\ handling. 
Here, tuning the full Morotti model \cite{Morotti2014}, which 
explicitly considers \pka\ and \camkii signaling, to reflect excitation-contraction coupling in rats may be appropriate.
		
\section*{Conclusions}
Our  predictions of elevated calcium transients under enhanced SL 
\catwo\ leak (via amylin oligomers) relative to control are in qualitative agreement with findings from Despa \etal\citep{Despa2012}. 
Further, these simulations suggest a potential mechanism linking human amylin infiltration of cardiac sarcolemma, amplification of intracellular \catwo\ transients and potential activation of \cam-dependent remodeling pathways; namely, amylin-induced increases in \SL\ \catwo\ leak potentially dually elevate  \catwo\ load in the cytosol and sarcoplasmic reticulum.
Increased sarcoplasmic reticulum \catwo\ content facilitates \catwo\ release, while elevated cytosolic \catwo\ levels promote the activation of \catwo-dependent proteins, including \cam.
The latter effect may potentially contribute to the \cam-dependent activation of NFAT/HDAC pathways reported in \citep{Despa2012}. 
Given that human amylin oligomers have been shown to deposit in  cell types including cardiac, neuronal, microglia, and beta cells \cite{Zhang2017,Haataja2008,Verma2016,Despa2012}, the effects of amylin-induced \catwo\ dysregulation may generalize to a variety of pathologies in higher animals.

\section*{Acknowledgments}
PKH thanks the University of Kentucky for pilot grant support, as well as a grant from the
National Institute of General Medical Science (P20 GM103527) of the National Institutes of Health. 
This	work	was	also	supported	by	the	National	Institutes	of	Health	(R01HL118474	to	FD	and	R01HL109501	to	SD)	and	The National	Science	Foundation	(CBET	1357600	to	FD).

\clearpage
\newpage 


\bibliography{pkh_sub}       
\clearpage
\newpage
\section*{Supplement}
\setcounter{figure}{0}
\makeatletter
\renewcommand{\thefigure}{S\@arabic\c@figure}
\makeatother

\setcounter{table}{0}
\makeatletter
\renewcommand{\thetable}{S\@arabic\c@table}
\makeatother

\setcounter{equation}{0}
\makeatletter
\renewcommand{\theequation}{S\@arabic\c@equation}
\makeatother

\newcommand{\uauf}
{\SI{}{\micro \amp \per \micro \farad}}
\newcommand{\msmf}
{\SI{}{\milli \siemen \per \micro \farad}}
\newcommand{\nspf}
{\SI{}{\nano \siemen \per \pico \farad}}
\newcommand{\mMms}
{\SI{}{\milli \molar \per \milli \second}}
\newcommand{\lfms}
{\SI{}{\liter \per \farad \per \milli \second}}

\subsection*{Supplemental Results} 
\subsubsection*{Validation of murine SB model} 
\label{supp:validation}
\paragraph{Potassium channel equations}
Among the most significant changes in the murine-specific Morotti model relative to the Shannon-Bers rabbit ventricular myocyte system are the phenomological representations of \kp\ channel currents. 
In \fig{figshort:Kcurrents} we compare time-dependent current profiles for nine of the prominent \kp\ channels. 
Nearly all channels required minor parameter changes to correspond to murine species (\ikp,\inak,\iki,\ikr,\itof, \iks, and \itos, of which the latter two were inactive in mice); however, two channels, \iss\ and \ikur, were not included in the Shannon-Bers model and are thus implemented here.
Following Morotti et al. \citep{Morotti2014}, \iss\ (\eqn{eq:i_ss}) and \ikur\ (\eqn{eq:i_Kur_PKAp}) were parameterized as follows: 

\textbf{IKur}\\
\label{comp:IKur}
\begin{dgroup}
  \begin{dmath}
    \label{eq:i_Kur_PKAp}
    i_{Kur PKAp} = \frac{y_{174}}{i_{KurtotBA}}\\
  \end{dmath}
  \begin{dmath}
    \label{eq:a_Kur}
    a_{Kur} = \frac{0.2}{-1 + \frac{fIKurpISO}{fIKurp_{0}}}\\
  \end{dmath}
  \begin{dmath}
    fIKuravail = 1 - a_{Kur} + \frac{a_{Kur}}{fIKurp_{0}} i_{Kur PKAp}\\
  \end{dmath}
  \begin{dmath}
    \label{eq:fIKuravail}
    fIKuravail = 1\\
  \end{dmath}
  \begin{dmath}
    \label{eq:i_kur1}
    i_{kur1} = Gkur_{1} Kcoeff fIKuravail \left(- E_{K} + V\right) X_{Kur slow} Y_{Kur slow1}\\
  \end{dmath}
  \begin{dmath}
    \label{eq:i_kur2}
    i_{kur2} = Gkur_{2} Kcoeff \left(- E_{K} + V\right) X_{Kur slow} Y_{Kur slow2}\\
  \end{dmath}
  \begin{dmath}
    \label{eq:i_kur}
    i_{kur} = i_{kur1} + i_{kur2}\\
  \end{dmath}
\end{dgroup}
\textbf{Xkur\_gate}\\
\label{comp:Xkur_gate}
\begin{dgroup}
  \begin{dmath}
    \label{eq:X_Kur_slowss}
    X_{Kur slowss} = \frac{1}{1 + 0.34 e^{- 71.4\!\times\!10 ^{-3} V}}\\
  \end{dmath}
  \begin{dmath}
    \label{eq:tau_Xkur}
    \tau_{Xkur} = 0.95 + 50\!\times\!10 ^{-3} e^{- 80\!\times\!10 ^{-3} V}\\
  \end{dmath}
  \begin{dmath}
    \label{eq:dX_Kur_slow_dt}
    \frac{dX_{Kur slow}}{dt} = \frac{1}{\tau_{Xkur}} \left(- X_{Kur slow} + X_{Kur slowss}\right)\\
  \end{dmath}
\end{dgroup}
\textbf{Ykur\_gate}\\
\label{comp:Ykur_gate}
\begin{dgroup}
  \begin{dmath}
    \label{eq:Y_Kur_slowss}
    Y_{Kur slowss} = \frac{1}{1 + 2.3\!\times\!10 ^{3} e^{0.16 V}}\\
  \end{dmath}
  \begin{dmath}
    \label{eq:tau_Ykur1}
    \tau_{Ykur1} = 400 - \frac{250}{1 + 553\!\times\!10 ^{-6} e^{- 0.12 V}} + 900 e^{- \left(3.44 + 62.5\!\times\!10 ^{-3} V\right)^{2}}\\
  \end{dmath}
  \begin{dmath}
    \label{eq:dY_Kur_slow1_dt}
    \frac{dY_{Kur slow1}}{dt} = \frac{1}{\tau_{Ykur1}} \left(- Y_{Kur slow1} + Y_{Kur slowss}\right)\\
  \end{dmath}
  \begin{dmath}
    \label{eq:tau_Ykur2}
    \tau_{Ykur2} = 400 + \frac{550}{1 + 553\!\times\!10 ^{-6} e^{- 0.12 V}} + 900 e^{- \left(3.44 + 62.5\!\times\!10 ^{-3} V\right)^{2}}\\
  \end{dmath}
  \begin{dmath}
    \label{eq:dY_Kur_slow2_dt}
    \frac{dY_{Kur slow2}}{dt} = \frac{1}{\tau_{Ykur2}} \left(- Y_{Kur slow2} + Y_{Kur slowss}\right)\\
  \end{dmath}
\end{dgroup}
\textbf{Xss\_gate}\\
\label{comp:Xss_gate}
\begin{dgroup}
  \begin{dmath}
    \label{eq:xssss}
    xssss = X_{Kur slowss}\\
  \end{dmath}
  \begin{dmath}
    \label{eq:tauxss}
    tauxss = 14 + 70 e^{- \left(1.43 + 33.3\!\times\!10 ^{-3} V\right)^{2}}\\
  \end{dmath}
  \begin{dmath}
    \label{eq:dXss_dt}
    \frac{dXss}{dt} = \frac{1}{tauxss} \left(- Xss + xssss\right)\\
  \end{dmath}
  \begin{dmath}
    \label{eq:i_ss}
    i_{ss} = Gss Kcoeff \left(- E_{K} + V\right) Xss\\
  \end{dmath}
\end{dgroup}

In \fig{figshort:rabbitvsmouse}, we summarize several key predicted outputs for the Shannon-Bers (rabbit) and \sbm (mouse) models: cytosolic  and \sr\ \catwo\ transients (a-b), as well as sodium load (c) and action potential (d). 
The \sbm\ implementation exhibits, for instance, modestly higher \catwo\ transients in both the cytosol and \sr, higher intracellular \nap\ load and a significantly shorter \ap, in comparison to data predicted for rabbit \catwo\ handling.
The decreased action potential largely stems from reparameterization of the potassium currents defined above, the currents of which we summarize in \fig{figshort:Kcurrents}.
These model predictions are in quantitative agreement with corresponding current profiles presented in the Morotti \etal\ supplemental data \citep{Morotti2014}, which were based on transient data from Dybkova \etal \citep{Dybkova2011}. 
We present analogous current data for \ina, \ilcc, and \incx, which again are in quantitative agreement with Morotti \etal. 
Altogether, these predictions indicate that our implementation of the Morotti model faithfully reproduces the murine electrophysiology and \catwo\ handling. 
This implementation serves as the basis for our further refinement to reflect the rat \catwo\ dynamics.

\subsubsection*{Maintenance of sodium load}
In the context of \catwo\ handling, \nap\ serves an important role in both extruding cytosolic \catwo\ in its 'forward' mode, as well as promoting \catwo\ influx during its brief 'reverse' mode \citep{Bers2001}.
Sodium load exceeding normal physiological ranges (approximately 9-14 mM in rodents), for instance, can contribute to diastolic dysfunction \citep{Louch2010a,Despa2013}, predominantly by impairing \ncx\ \catwo\ extrusion \cite{Altamirano2006}.
Conversely, the \ncx\ reverse mode may leverage \nap\ gradients to amplify \sl\ \catwo\ transients and thereby prime SR \catwo\ release \citep{Swift2008,Litwin1998,Sipido1997a,Sobie2008,Kekenes-Huskey2012}.
While our measurements of \nap-load in \amy\ rats indicated that intracellular \nap\ was within normal ranges, numerical predictions suggested that loading may be elevated under conditions of increased \sl\ \catwo\ leak with constant \nka\ function.
Therefore to maintain predicted \nap\ transients within control levels, a modest increase in \nka\ Vmax was predicted. 
On one hand, there is precedent for small peptides like insulin partitioning into the rat skeletal transverse tubule system \cite{Shorten2007}, as well agonism of \nka\ activity due to amylin \cite{Clausen2004}.
However, for rat cardiac ventricular tissues, these changes in \nka\ function may be non-existent or below the limits of experimental detection at least in fully-developed diabetes \citep{Lambert2015}. 

\subsubsection*{Up-regulated SL currents (via LCC) have dissimilar phenotype to amylin action}
The dominant effect of amylin appears to be its enhancement of non-selective SL \catwo\ currents, although there are reports that amyloidogenic peptides can alter \lcc\ regulation \citep{Kim2011}.
To delineate this contribution from secondary agonism of SL \catwo\ channel activity, we performed simulations using an amplified \ilcc\ sufficient to reproduce the \catwo\ transients observed for enhanced SL \catwo\ leak. 
We emphasize here that LCC current/voltage relationship is indeed preserved in \hip\ rats (as shown in \fig{figshort:lccexp}).
Similar to \amy, increased LCC current yielded increased intracellular \catwo\ transients, elevated SR \catwo\ load, and increased diastolic \catwo\ load.
Nevertheless, we identified distinct patterns of modulated channel activity for increased LCC relative to those presented for the \amy\ configuration.
Namely, our models indicate amylin-induced \catwo\ leak amplifies \ina, whereas in contrast, increased LCC conductance inflates the amplitudes of several prominent \kp\ channels.
While the predicted channel currents have a complex dependence on ion-sensitive gating probabilities, these findings raise interesting possibilities that different modes of \catwo\ entry could in principle yield distinct influences on channels controlling the action potential.
Nevertheless, under the conditions considered in this study, our modeling data (see \fig{figshort:amy} and \fig{figshort:hip}) suggest that the modest perturbations in ion channel conductance under the \amy\ and \lcc\ configurations did not appreciably impact the \apd. 
These findings are consistent with preserved \apd\ upon P2X stimulation reported in Fig 6 of \citep{Shen2007}.

\subsubsection*{Genetic algorithm for fitting}
\label{supp:fitting}
In order to optimize model fitting to the experimentally measured cases, a genetic algorithm was written to fit the model to various cases.
For example, in regards to the Amylin case, we randomized the \nka\ current parameter to find what value that gave the closest steady state value of 12 mM for the intracellular \nap\ over 30s of simulation time.
A starting value for \nka\ current (5 \uauf) was given with a generous starting standard deviation (0.5) to randomize within.
The standard deviation value was done as a lognormal distribution to ensure that reductions and increases of N\% \mynote{PKH: I am a little confused here on the purpose of the lognormal distribution} were equally probable.
A total of 30 random draws were made.
Once the random draws were made, they were submitted to be run using CPUs.
After the job finished, an error value was calculated for comparing an output of the system (in this case intracellular \nap) against the experimentally determined value by squaring the result minus the experiment value.

This gave a "job fitness" score that was used to determine which random draw was the best. 
\begin{align}
	jobFitness = (X_{i,exp}-X_{i,truth})^{2}
\end{align}
The job that had the best "job fitness" score was now selected as the new starting value to have jobs randomized around for the next iteration.
The standard deviation was then adjusted by multiplying the input standard deviation by e to the negative iteration number\bdsnote{why? (explain)} times a scaler. 
\begin{align}
	\sigma_{i} = \sigma_{0} e^{-i \sigma_{scaler}}
\end{align}
This new standard deviation determined the range of random draws around the new starting value given from the first iteration of the code.
This process was then continued for several different iterations (20) to converge to a single value of the \nka\ current which would also give an output value for intracellular \nap\ to match that of the experimentally given value. \bdsnote{give tolerance value used for convergence PKH: We didn't really have a tolerance value to match against.}
An example plot of the convergence of the \nka\ current value over number of iterations can be seen in \fig{figshort:algVerify}. \bdsnote{explain briefly what the data suggest and why we think it's working DONE}
The graph shows how after the amount of iterations increased, the value of the \nka\ current converged. 
This value of the \nka\ current was then used within the simulation and the intracellular \nap\ value was compared to the baseline case to match experiment. 
This plot can be see in \fig{figshort:amy}.
Since the value found for the \nka\ current using the genetic algorithm gave an intracellular \nap\ comparable to that of experiment, it can be seen that the generic algorithm worked as expected.

\subsubsection*{Comparison of parameter sensitivity}
\paragraph*{Sensitivity analyses}
\label{supp:sensanaly}
We determined the sensitivity of \sbm\ model outputs including \catwo\ amplitude, cytosolic \nap, \sr\ \catwo, diastolic \catwo, \apd, and \catwo\ transient decay ($\tau$) 
to the model parameters, by randomizing model parameters temperature, background \catwo leak, background \nap\ leak, \serca\ function, \nka\ function, and PCa.
Each parameter was randomized independently, while holding all other parameters at their default values for the rat model. 

The random draw was done within a standard deviation value of 10\% of the given input value based on the baseline rat data.
This was done for a total of 10 times to get 10 random draws for the parameter. 
Once the 10 random draws were done for the chosen parameter, the program moved to a new parameter and repeated the process.
This lead to a total of 60 jobs that were run. 
Descriptive statistics (mean (M), standard deviation (SD), median (med)) and plots (boxplots, scatterplots with LOESS curves) were initially performed to assess distributional characteristics and bivariate relationships.
Multivariate multiple regression (Rencher \& Christensen, 2012)\bdsnote{put into mendeley} was performed to assess the simultaneous effects of each input on all outputs jointly. 
If multivariate Wilks’ $\Lambda$ tests for any associations with any output were significant, then tests for associations for individual outputs were performed in a step-down fashion. 
All analyses were performed in SAS v9.4 (SAS Institute, Cary, NC). 
A two-sided p-value $<$ 0.05 was considered statistically significant.

\paragraph*{Sensitivity results}
\label{supp:results}
Inputs that had at least one significant association with any of the outputs included: Background \catwo leak \bdsnote{replace variable names with what they mean (e.g GCaBk s.b. sarcolemmal calcium leak conductance etc) DONE} ($\Lambda$ = 0.039, F(5,49) = 242.804, $\eta^{2}$ = 0.961), background \nap leak ($\Lambda$ = 0.009, F(5,49) = 1,097.591, $\eta^{2}$ = 0.991), \nka\ Vmax ($\Lambda$ = 0.001, F(5,49) = 12,009.757, $\eta^{2}$ = 0.999), PCa ($\Lambda$ = 0.002, F(5,49) = 5,906.809, $\eta^{2}$ = 0.998), T ($\Lambda$ = 0.002, F(5,49) = 5,145.939, $\eta^{2}$ = 0.998), and \serca\ Vmax ($\Lambda$ = 0.014, F(5,49) = 708.322, $\eta^{2}$ = 0.986). 
Thus, inputs respectively accounted for 96\% or more variation in the best linear combination of outputs (all $\eta^{2}$ $\geqq$ 0.96). Specifically for increasing Background \catwo leak and increasing background \nap leak, there was only a significant association with increasing Nai (partial $\eta^{2}$ ($\eta^{2}_{p}$) = 0.568, p $<$ 0.001 for CaBk and $\eta^{2}_{p}$ = 0.829, p $<$ 0.001 for NaBk, respectively). 
Similar results were found for \nka\ Vmax. 
Here, decreasing Nai was the only output that was significantly associated with increasing \mnote{double check with TnT, since Ca transient changes seemed significant by eye} \nka\ Vmax ($\eta^{2}_{p}$ = 0.981, p $<$ 0.001). 
Increased PCa was significantly associated with both increased Nai ($\eta^{2}_{p}$ = 0.929, p $<$ 0.001) and increased APD ($\eta^{2}_{p}$ = 0.694, p $<$ 0.001). 
For T, higher Nai again was the only output significantly predicted by increasing T ($\eta^{2}_{p}$ = 0.951, p $<$ 0.001), while T was also marginally significantly associated with \sr\ \catwo\ ($\eta^{2}_{p}$ = 0.058, p = 0.076). 
Increased \serca\ Vmax was significantly associated with increased APD ($\eta^{2}_{p}$ = 0.166, p = 0.002), increased \sr\ \catwo\ ($\eta^{2}_{p}$ = 0.138, p = 0.005), and decreased Nai ($\eta^{2}_{p}$ = 0.354, p $<$ 0.001).

\clearpage
\newpage
\subsection*{Supplemental Tables} 
\label{supp:morottimodel}
\begin{sidewaystable}[h]
\centering
\caption{Comparison of \sb\ default parameters with mouse-specific variations in accordance to Morotti et al. \cite{Morotti2014} Places with * represent the value was fitted using our \ga\ to match corresponding experimental data.
\label{tbl:morottiparams}}
\begin{tabular}{lllll}
Parameter [units] & Name & Rabbit & Mouse & Rat \\
\hline
Membrane capacitance [F]& $C_m$ & \num{1.381e-10} & \num{2.0e-10} & - \\ 
Cell volume [L]    & $V_{Cell}$ & \num{30.4e-12}  & \num{33.0e-12}\citep{Morotti2014} & - \\ 
Intracellular \nap [mM] & $Na_i$ & 8.80853  & 11.1823   & 12.0 \\
Free concentration of \catwo\  & $[Ca]_o$ & 1.8 & 1.0 &  - \\ 
in extracellular compartment [mM] \\
Maximal conductance of flux for & $G_{INa}$ & 16 & 10 &  - \\ 
fast \nap\ current [\msmf] \\
Junctional partitioning & $F_{Xi}$ junction & 0.11 & 0.19 & - \\ 
Sarcolemmal partitioning& $F_{Xi}$ \sl & 0.89 & 0.81 & - \\ 
Background \nap\ leak [\msmf] & $G_{NaBk}$ & \num{0.000297} & \num{0.001337} &  - \\ 
K_{m} of \nap\ - \catwo\ exchanger for \nap\ [mM] & $K_{mNai}$ & 11 & 19 &- \\ 
Max current of the \nap\ - \kp\ pump [\uauf] & $I_{NaKmax}$ & 1.90719 & 5.0 & * \\ 
Slow activating delayed rectifier current  & $G_{Ks}$ & 1.0 & 0.0 & - \\ 
Slow inactivating delayed rectifier current 1 [\nspf] & $Gkur1$ & N/A & 0.176 &- \\ 
Slow inactivating delayed rectifier current 2 [\nspf] & $Gkur2$ & N/A & 0.14 &- \\ 
\iss conductance [\nspf] & $Gss$ & N/A & 0.15 &- \\ 
Velocity max for  \nap\ - \catwo\ exchanger [\uauf] & $V_{maxINaCa}$ & 9 & 1 & - \\ 
Allosteric \catwo\ activation constant [mM]& $K_{d-Act}$ & \num{0.000256} & \num{0.000128} &- \\ 
Background \catwo\ leak  [\msmf] & $G_{CaBk}$ & \num{0.0002513} & \num{0.0007539} & - \\ 
\sr\ \catwo\ concentration dependent\\ activation of \sr\ \catwo\ release  [mM] & $EC_{50-SR}$ & 0.45 & 0.5 &- \\ 
Passive leak in the \sr\ membrane  [\SI{}{\per \milli \second}] & $K_{SRleak}$ & \num{5.348e-6} & \num{10.7e-6} & - \\ 
Velocity max for \sr\ \catwo\ pump flux   [\mMms] & $V_{maxJpump}$ & \num{5.3114e-3} & * & * \\ 
$K_{m}$ \sr\ \catwo\ pump forward mode  [mM] & $K_{mf}$ & \num{0.000246} & \num{0.0003} & - \\ 
$K_{m}$ \sr\ \catwo\ pump reverse mode  [mM] & $K_{mr}$ & 1.7 & 2.1 &- \\ 
L-type \catwo\ channel \catwo\ permeability [\lfms] & $P_{Ca}$ & \num{5.4e-4} & \num{8.91e-4} & - \\ 
L-type \catwo\ channel \nap\ permeability  [\lfms] & $P_{Na}$ & \num{1.5e-8} & \num{2.475e-8} & - \\ 
L-type \catwo\ channel \kp\ permeability [\lfms] & $P_{K}$ & \num{2.7e-7} & \num{4.455e-7} &- \\ 
Background \catwo\ leak [\msmf] & $G_{CaBk}$ & \num{0.0002513} & \num{0.0007539} & - \\ 
\hline
\end{tabular}
\end{sidewaystable}

\begin{table}[]
\centering
\begin{tabular}{@{}ll@{}}
\toprule
Model term & Description \\\midrule
\iclca &         \catwo-activated chloride current \\
\iclb &          Background \ch{Cl} current\\
\icap &          \SL-\catwo\ pump \\
\icab &          Background \catwo\ leak\\
\inab &          Background \nap\ leak\\
\inak &          \nka\ current\\
\itof &          fast Cardiac transient outward potassium\\
\itos &          slow Cardiac transient outward potassium\\
\ikr &           the 'rapid' delayed rectifier current \\
\iks &           slowly activating K+ current\\
\iki &           inward rectifier K+ current\\
\ikp &           plateau potassium current\\
\ikur &          slowly inactivating outward \\
\iss &            non-inactivating steady-state K+ current\\
\ilcc &          \lcc\ channel current\\
\incx &          \ncx\ current\\
\ina &           \nap\ current\\
\slna &          \SL\ \nap\ \\
\jctna &         JSR \nap\ \\
I & \ryr\ inactive gate\\
O & \ryr\ open gate \\
Cai & cytosolic \catwo\  \\
V & Action potential \\
\bottomrule
\end{tabular}
\caption{Model terms. }
\label{tbl:modelterms}
\end{table}

\begin{sidewaystable}
\centering
\caption{Parameters used in \sbm\ computational model to reflect  control and hyperamylinemia/pre-diabetic rat. 
Percentages in parentheses are relative to control rat.
$^{I}$from \citep{Despa2012}. 
$^{II}$from \citep{Morotti2014}. 
$^{III}$fitted to \citep{Despa2012}. 
$^{IV}$fitted to \citep{Gattoni2016}.
\label{tbl:params}}
\begin{tabular}{lcccc}
Case  & Sarcolemmal leak  & SERCA            & NKA & PCa \\
      & $G_{Ca}$   & $V_{max}$ & $I_{NKA\_max}$  & \\
      & (mS/$\mu$F)       & (mM/ms)          & ($\mu$A/$\mu$F)          &\\
\hline
control &  $^{II}$\num{0.00075}   &  $^{IV}$\num{0.010}  & $^{III}$3.85\\
\amy\   &  $^{III}$\num{0.0013}, (168\%)  & $^{III}$\num{0.010}  & $^{III}$4.40, (114\%)\\
\hip\   &   $^{III}$\num{0.0013}, (168\%) &  $^{III}$\num{0.0053},(53\%) \ & $^{III}$4.40, (114\%)\\
LCC     &  $^{II}$\num{0.00075} & $^{IV}$\num{0.010} & $^{III}$3.85 &  $^{III}$\num{0.0016038},(180\%) \\
\hline                    
\end{tabular}
\end{sidewaystable}

\clearpage
\newpage
\subsection*{Supplemental Figures} 
\figbig{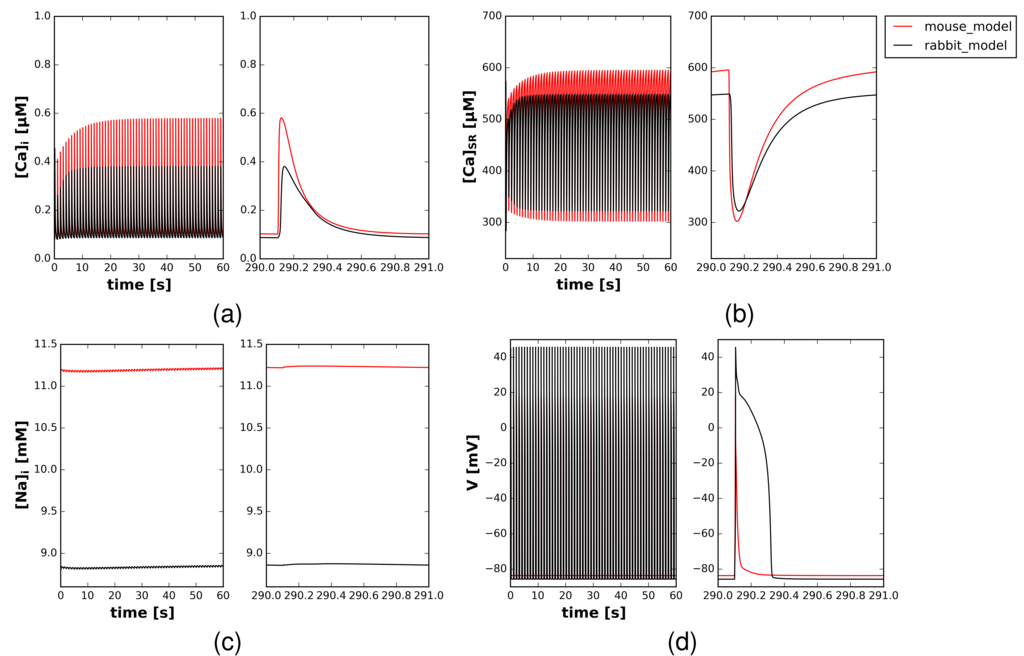}
{rabbitvsmouse}
{Predicted intracellular \catwo\ (a), sarcoplasmic reticulum \catwo\ (b), intracellular sodium (c), and action potential (d) for mouse (black) and rabbit (red) conditions}
{morottiModel_daisychain.ipynb}

\figbig{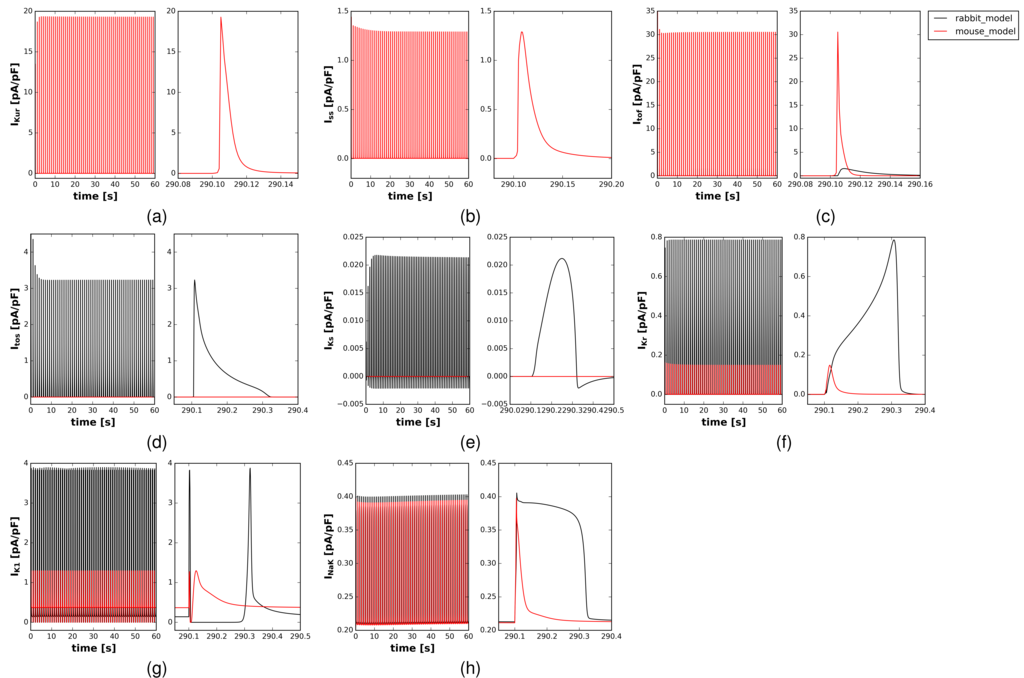}
{Kcurrents}
{Comparison of potassium currents predicted for rabbit (black) and mouse (blue) ventricular cardiomyocytes via the \sb\ and \sbm\ models, respectively. Top row, from left: slowly inactivating current, \ikur, steady-state current, \iss, fast transient outward current, \itof. Middle row: slow transient outward current, \itos, slowly activating current, \iks, rapidly activating current, \ikr. Bottom row: inward rectifier current, \iki, sodium/potassium exchanger, \inak} 
{morottiModel_daisychain.ipynb}
\newpage

\figbig{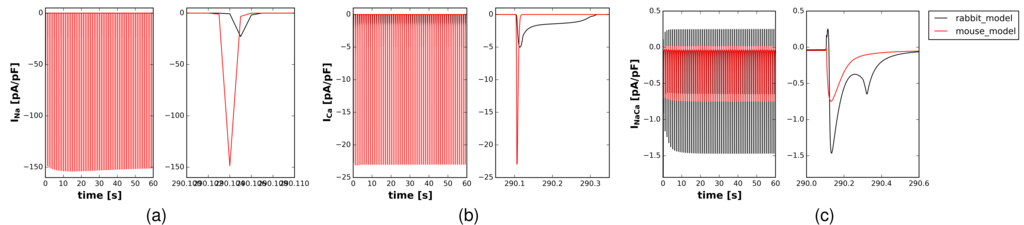}
{othercurrents}
{Comparison of sodium current (Left), \ina, L-type \catwo\ channel current (middle), \ilcc, and sodium/\catwo -exchanger current (right), \incx,  predicted for rabbit (black) and mouse (blue) ventricular cardiomyocytes via the \sb\ and \sbm\ models, respectively. \mynote{BDS:please run at higher temporal resolution (5-10x smaller dt) PKH: code is hardcoded to downsample right now. will try to change this when I can.}}
{morottiModel_daisychain.ipynb}

\figbig{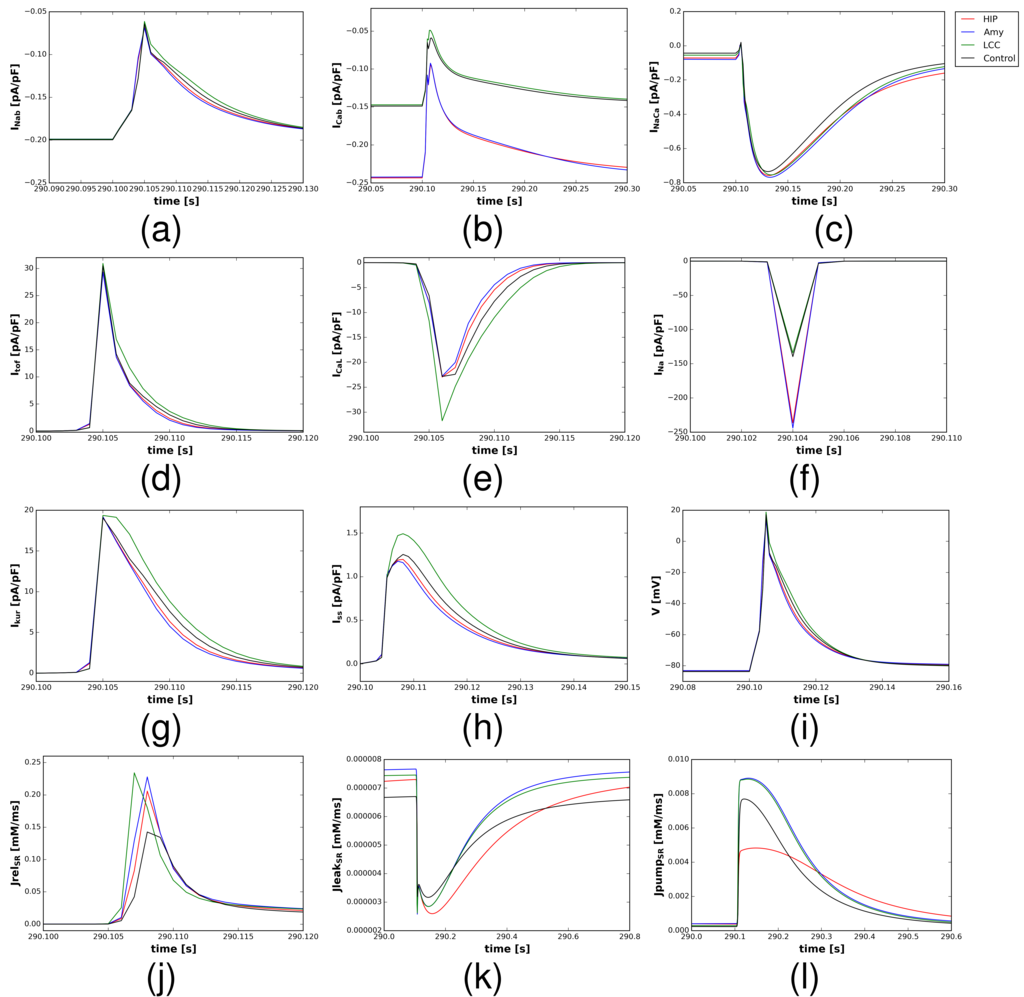}
{Cabsetc}
{Fluxes and currents for control (black), \amy (blue), \hip (red), and LCC (green) conditions
\mynote{BDS:please replace panel i with a single AP waveform DONE}
}
{morottiModel_daisychain.ipynb}

\figbig{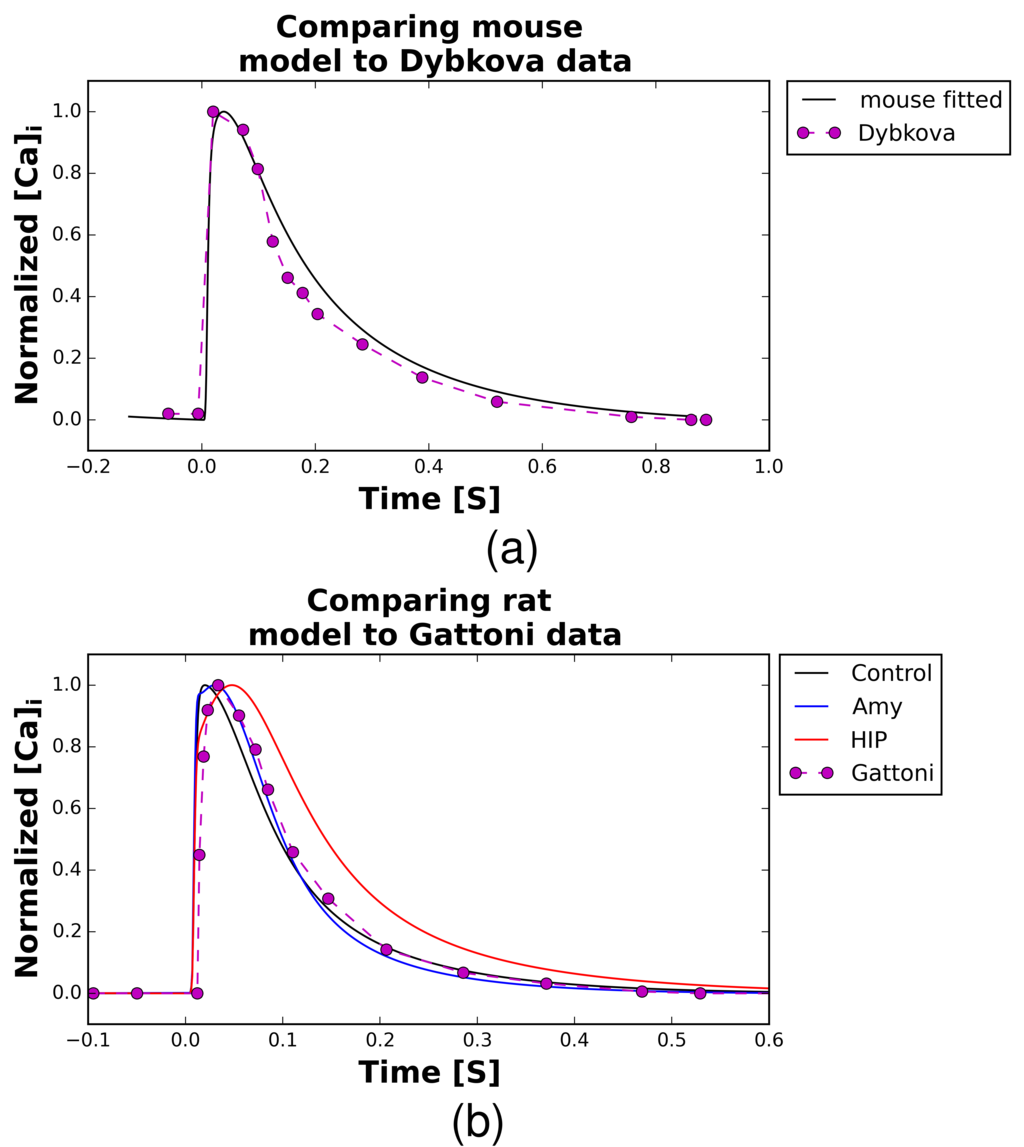}
{expfitting}
{
Comparison of experimental \catwo\ transient data at 1 Hz for mouse (a)\citep{Dybkova2011,Morotti2014} and rat (b)\citep{Gattoni2016} (purple) with our predicted control data (black) . \catwo\ transients predicted for \amy\ (blue) and \hip\ (red) are additionally provided.}
{morottiModel_daisychain.ipynb}

\figbig{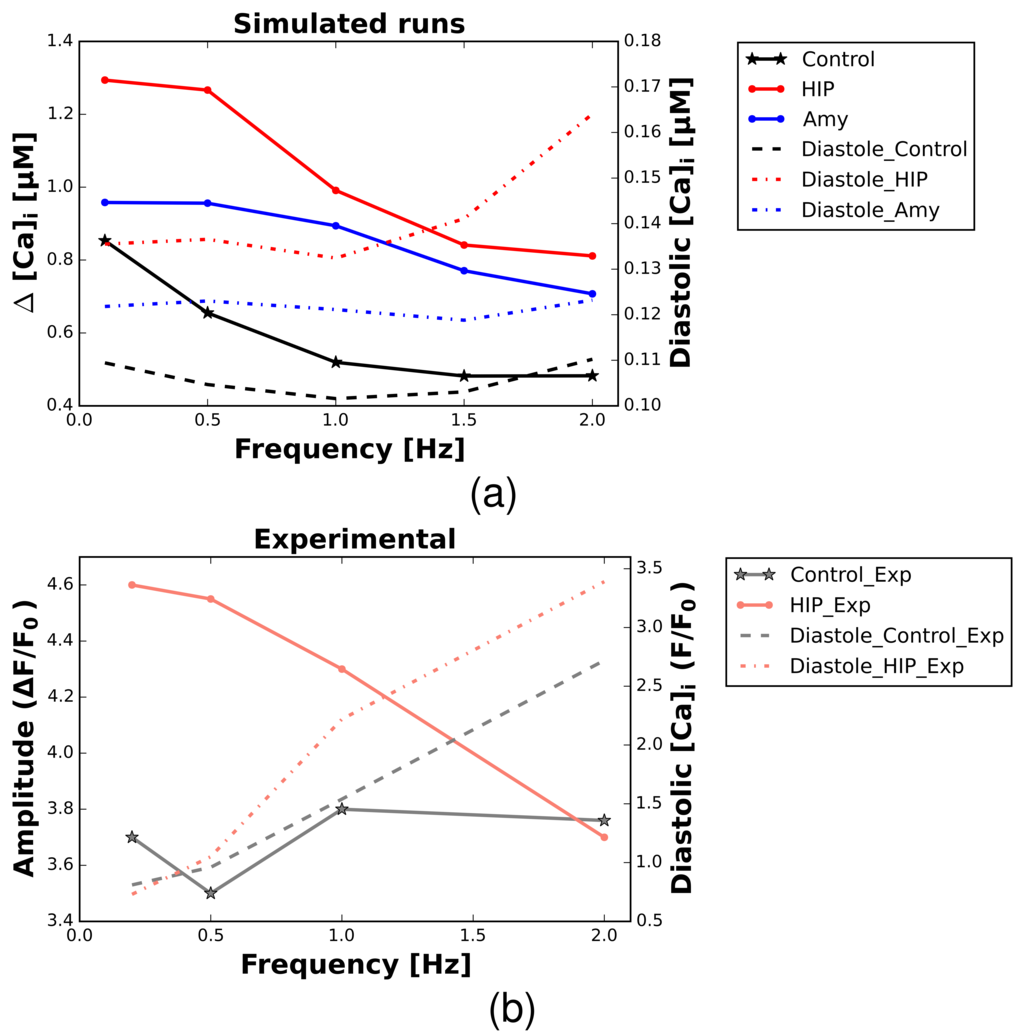}
{ratfigure5D}
{Predicted intracellular \catwo\ transient amplitude ($\Delta Ca$ [uM], left axis,solid) and diastolic \catwo\ load (right axis, dashed) versus pacing frequency [Hz] for control (black) and HIP (red) conditions. 
Data are provided based on \sbm\ model predictions (a) and Despa et al \citep{Despa2012} (b).
}
{morottiModel_daisychain.ipynb}

\figbig{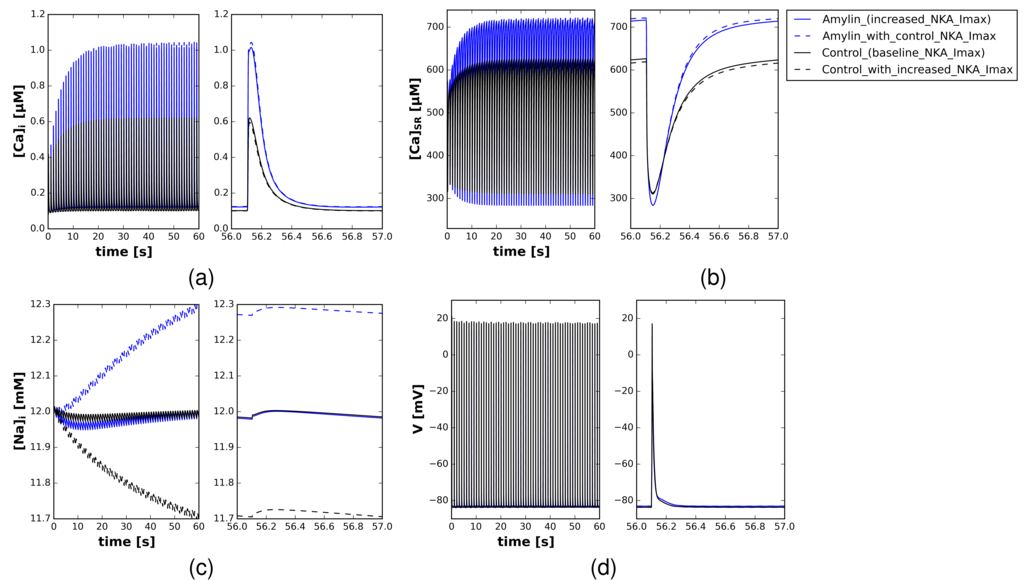}
{nka}
{Simulated contributions of \SL\ sodium/potassium ATPase activity to intracellular \catwo\ and \nap\ load. 
Predicted a) \catwo\ and b) \nap\ intracellular transients under 
control (black, solid line), 
control with increased NKA to match \amy\ level NKA current(black, dashed line), 
\amy\ (blue, solid line), and 
\amy\ with decreased NKA to match control level NKA current (blue, dashed line).
See \tbl{tbl:params} for parameters. 
\mynote{BDS: Update legend to reflect 'Control (baseline NKA Vmax)','Control with increased NKA Vmax', 'Amylin (increased NKA)', 'Amylin with baseline NKA Vmax' DONE }
\mynote{move to supplement?}} 
{morottiModel_daisychain.ipynb}

\figbig{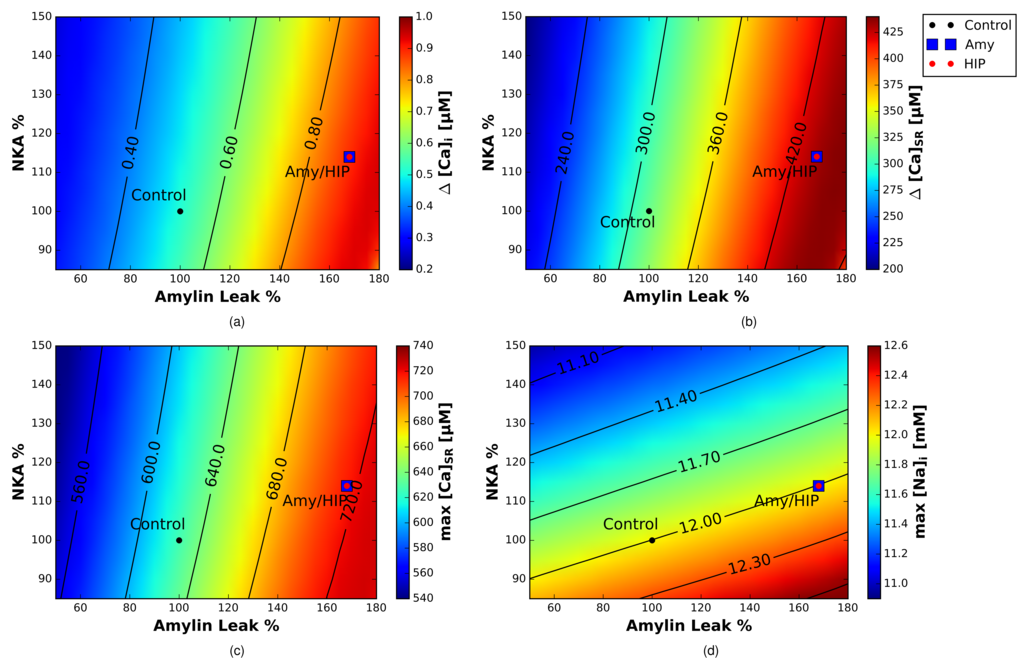}
{3dchartNKA}
{Predicted \catwo\ transients and loads as a function of NKA activity (\% of control) and SL \catwo\ leak (\% of control). a) intracellular \catwo, b) SR \catwo\ transient c) maximum SR \catwo\ load and d) sodium load. A black point is representative of the Control case, a blue square is representative of the Amy case, and a red point is representative of the HIP case.
Measurements are taken at 55 s}
{morottiModel_daisychain.ipynb}

\figbig{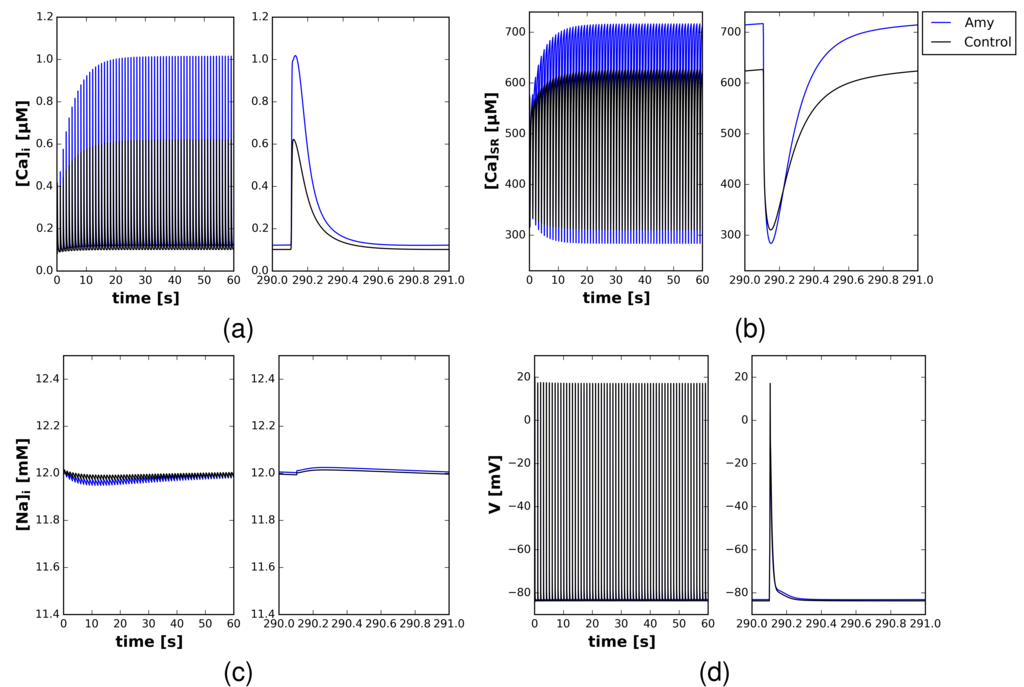}
{amy}
{Predicted intracellular \catwo\ (a), sarcoplasmic reticulum \catwo\ (b), intracellular sodium (c), and action potential (d) for control (black) and \amy\ (blue) conditions. Results are presented for 0 to 60s for clarity, although action potentials for up to 300s are reported in \fig{figshort:APfull}}
{morottiModel_daisychain.ipynb}


\figbig{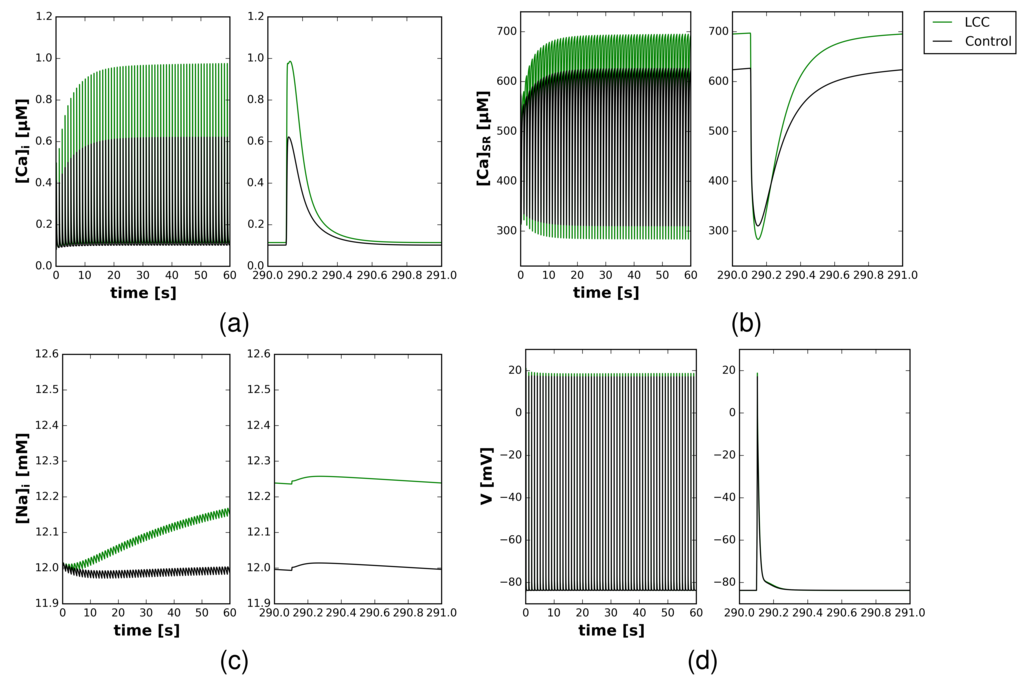}
{lcc}
{Predicted intracellular \catwo\ (a), sarcoplasmic reticulum \catwo\ (b), intracellular sodium (c), and action potential (d) for control (black) and increased \lcc\ current (green) conditions. Results are presented for 0 to 60s for clarity}
{morottiModel_daisychain.ipynb}

\figbig{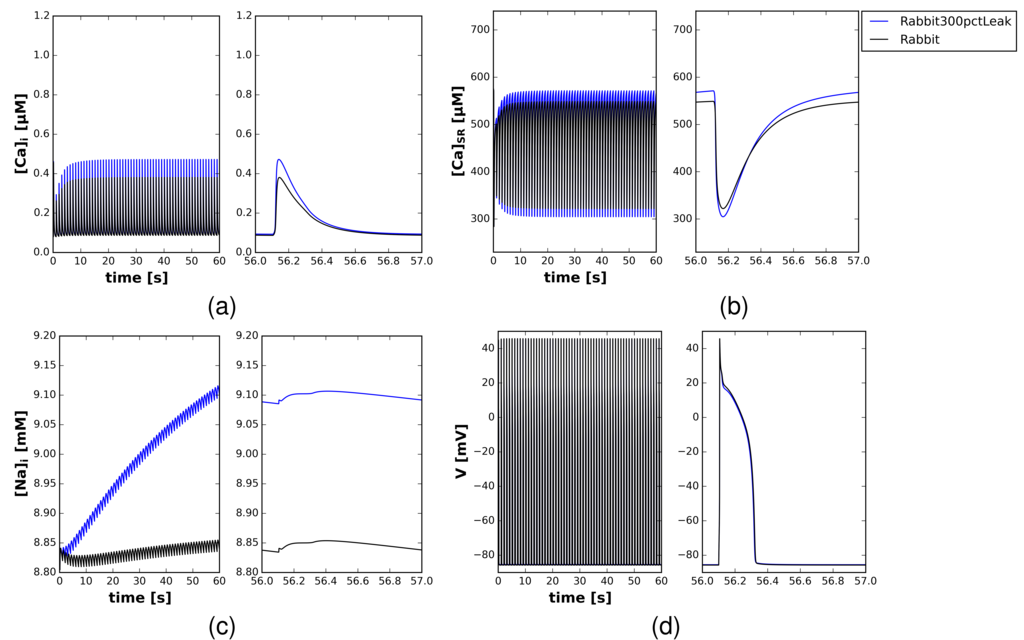}
{rabbitamyNoNKAleak300pct}
{Predicted intracellular \catwo\ (a), sarcoplasmic reticulum \catwo\ (b), intracellular sodium (c), and action potential (d) for control (black) and 300 \% increased \catwo\ background leak (blue) conditions using the rabbit \sb\ model.
\mynote{remove NoNKA from the label DONE}
}
{morottiModel_daisychain.ipynb}

\figbig{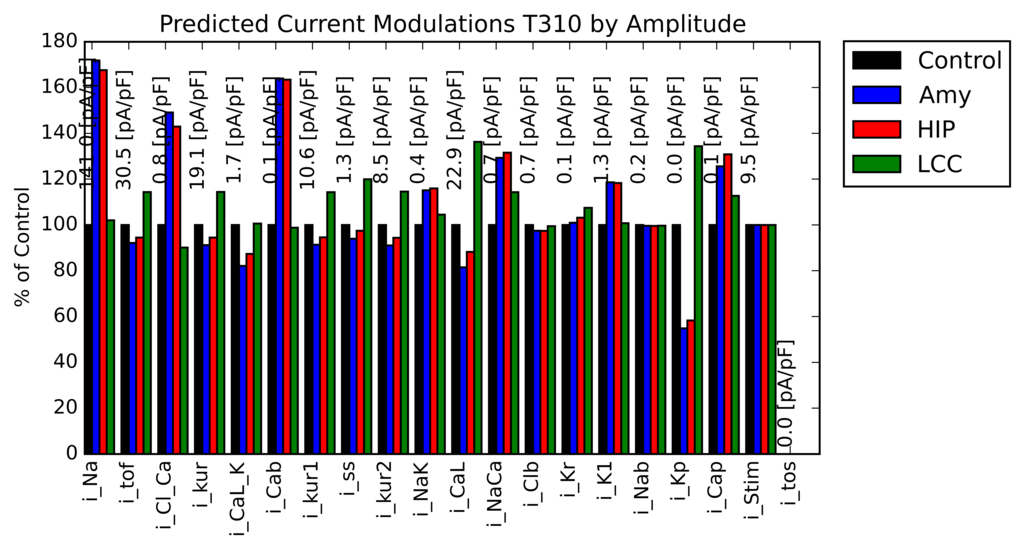}
{comparativecurrents}
{Percent of control (black) conditions for \sbm-predicted ion current amplitudes for \amy (blue), \hip (red) and increased LCC conductance (green, see \fig{figshort:lcc}) configurations. 
Currents are rank ordered by percent differences between control and \amy.  
The amplitude of the control current is provided as bar label. 
A list of current labels is provided in the supplement \tbl{tbl:modelterms}}
{morottiModel_daisychain.ipynb}

\figbig{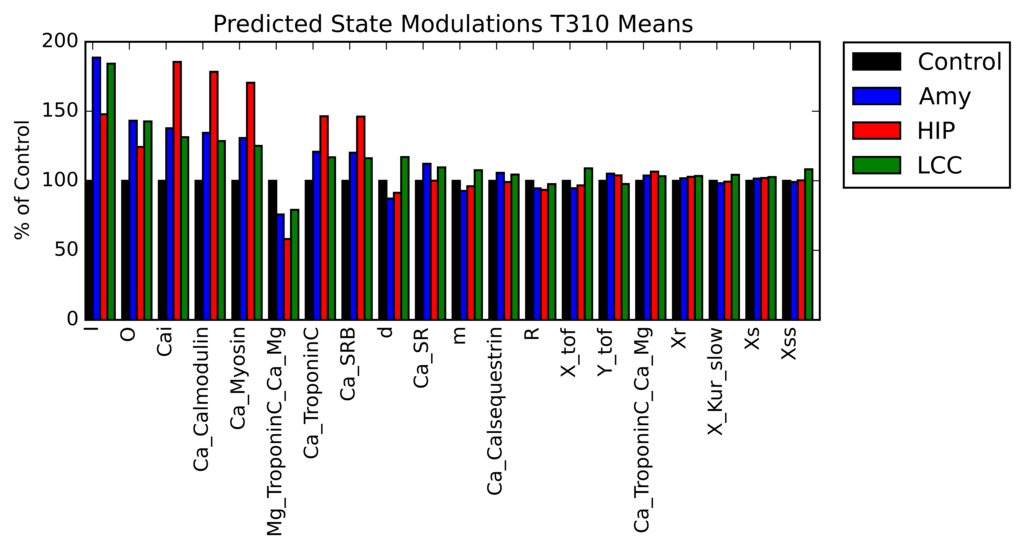}
{comparativestates}
{Percent of control (black) conditions for \sbm-predicted state variables, such as sarcolemma, cytosolic and junctional \catwo ($Ca_{SL}$,$Cai$, $Ca_{jct1}$) for \amy (blue), \hip (red) and increased LCC (green).
A list of state labels is provided in \tbl{tbl:modelterms}}
{morottiModel_daisychain.ipynb}

\figbig{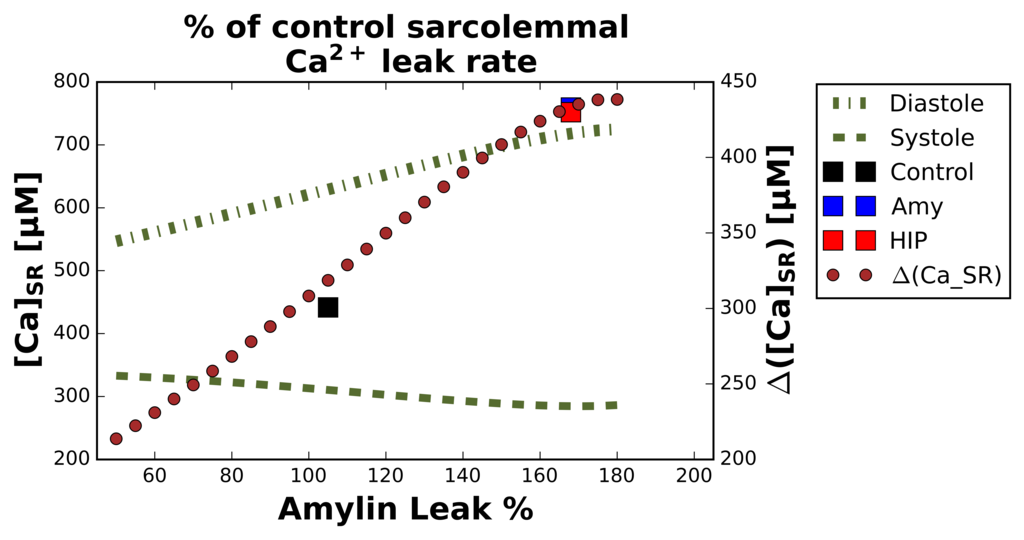}
{diffCaSRConc}
{Predicted sarcoplasmic reticulum \catwo\ load as a function of \acrfull{SL} \catwo\ leak rates (scaled relative to control) at 1 Hz pacing to approximate dose-dependent amylin incubation effects in rats.
Left axis: maximum \catwo\ (at diastole) and minimum (at systole, magenta dashed).
Right axis: \sr\ \catwo\ transient amplitudes (magenta dots) and squares indicating the \SL\ leak rates assumed for control (black), \amy\ (blue), and \hip\ (red) conditions}
{morottiModel_daisychain.ipynb}

\figbig{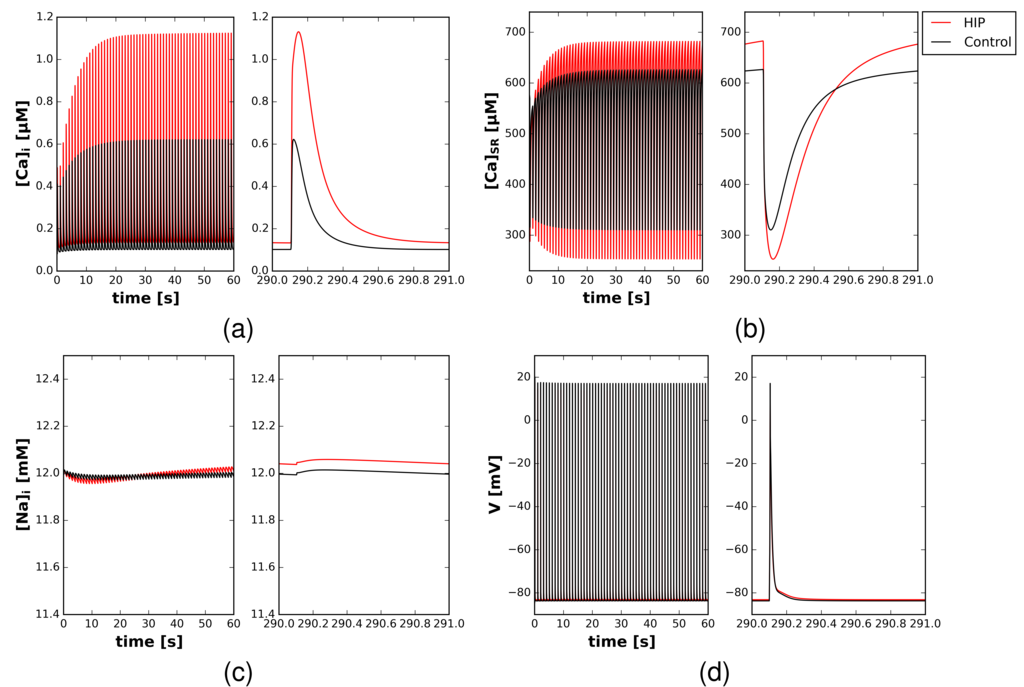}
{hip}
{Predicted intracellular \catwo\ (a), sarcoplasmic reticulum \catwo\ (b), intracellular sodium (c), and action potential (d) for control (black) and \hip\ (red) conditions. 
Results are presented for 0 to 60s for clarity, we report full 300s simulations in \fig{figshort:APfull}}
{morottiModel_daisychain.ipynb}

\figbig{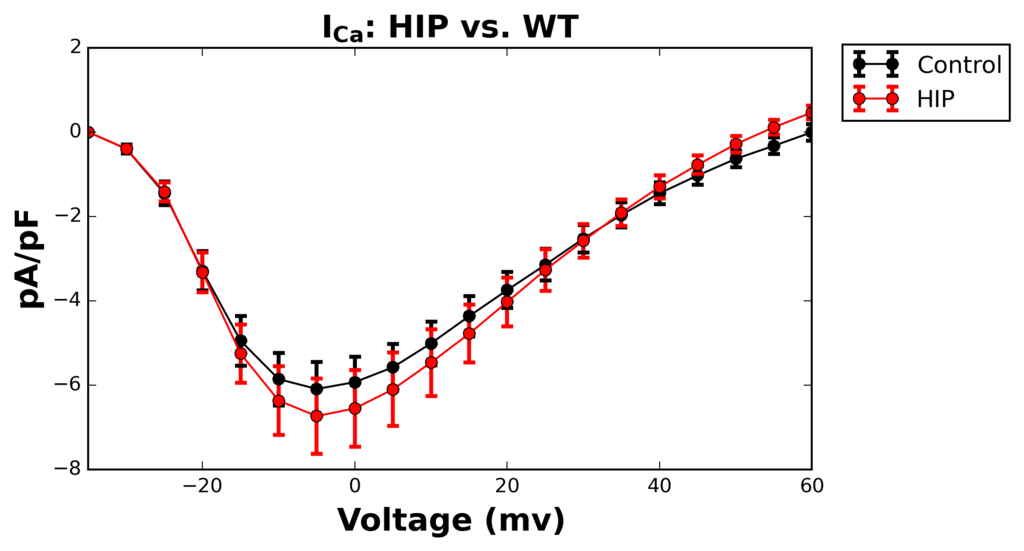}
{lccexp}
{Experimental recordings of current versus applied voltage for the \acrfull{lcc} in control (control) and \hip\ myocytes} 
{morottiModel_daisychain.ipynb} 

\figbig{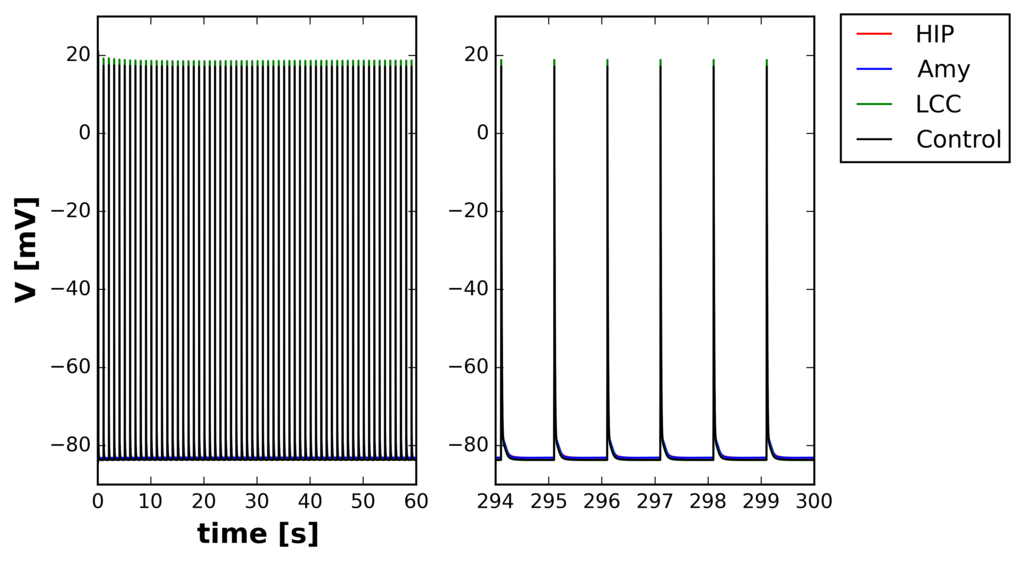}
{APfull}
{Predicted action potential, $V$, for control, \amy, \hip and LCC conditions over five minutes of 1 Hz pacing to demonstrate model stability
\mynote{BDS:Need to discuss downsampling technique used for data storage so that short-duration AP info is not lost}}
{morottiModel_daisychain.ipynb}

\figbig{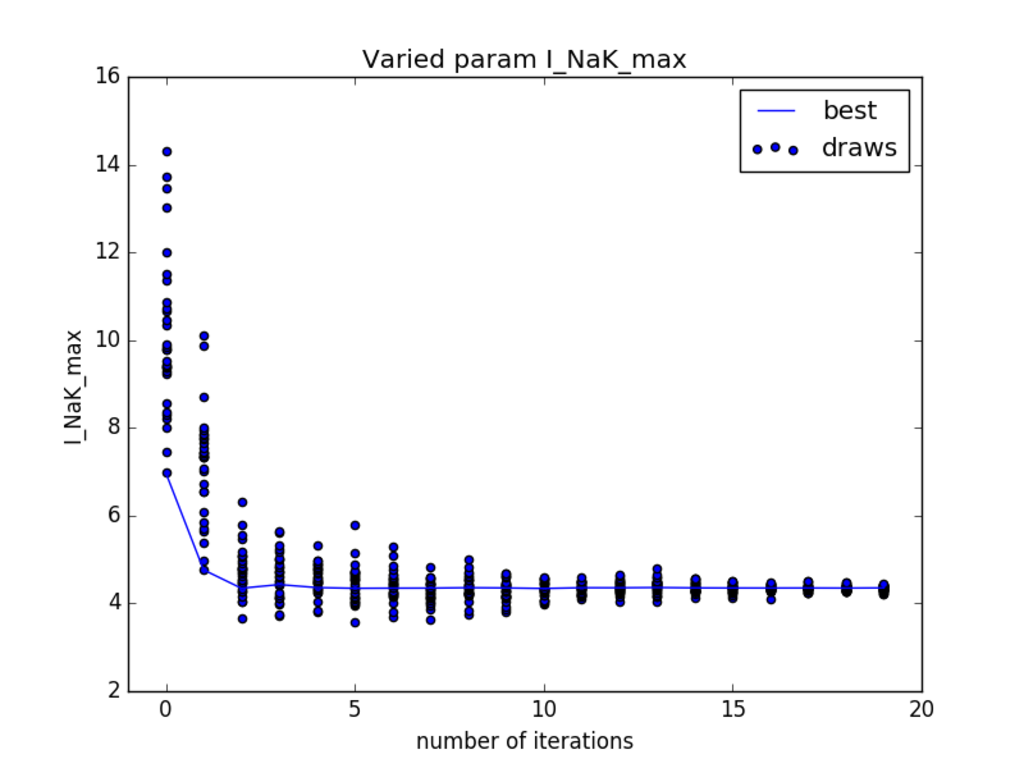}
{algVerify}
{Verification of the genetic algorithm used to fit the mouse model to the rat model. Many random draws were made for each iteration with the best random draw chosen. The best random draw becomes the new starting point random draws are made around for the next iteration. Each iteration was given a smaller range to randomize over in order to converge the system. As can be seen, the system converged to single value after completing several iterations} 
{FittingAlgorithmValidate}

\end{document}